\documentclass[showpacs,aps,prd,nofootinbib,floatfix,amsmath,amssymb]{revtex4}
\usepackage{graphicx}
\usepackage{dcolumn}
\usepackage{bm}
\usepackage[usenames]{color}
\usepackage{float}
\usepackage{pstricks}
\usepackage[english]{babel}
\usepackage[latin1]{inputenc}
\usepackage[usenames]{color}
\usepackage{hyperref}
\usepackage{subfigure}
\usepackage{caption}
\usepackage{setspace}

\newcommand{\be}{\begin{eqnarray}}
\newcommand{\ee}{\end{eqnarray}}
\newcommand{\nn}{\nonumber}
\begin{document}
\preprint{HEP-BUAP-01.13}

\title{Dipole moments of charged leptons in the THDM-III with Textures}

\author{M. Arroyo-Ure\~na and Enrique D\'iaz}
\affiliation{%
Facultad de Ciencias F\'isico-Matem\'aticas\\
Benem\'erita Universidad Aut\'onoma de Puebla, C.P. 72570, Puebla, Pue., Mexico.}%

\date{\today}

\begin{abstract}
We study the magnetic and weak magnetic dipole moments of charged leptons in the framework of the Two Higgs Doublet Model type III (THDM-III) with four zero textures. We first analyze the possibility that the discrepancy between the experimental measurements and the theoretical prediction of the muon anomalous magnetic dipole moment ($a_{\mu}$) in the Standard Model (SM) can be explained in the context of the model and then find values for the parameters of the THDM-III: neutral and charged Higgs boson masses ($m_{H}$, $m_{H^{\pm}}$), mixing angles ($\alpha, {\beta}$), CP-violating angles $\theta_1,\,\theta_2$ and an additional parameter arising from the mass matrix ($\gamma_f$), for which a correspondence between theory and experiment is possible. We use $a_{\mu}^{THDM-III}$ and current low energy processes $K-\bar{K}$ mixing, $B_s^0\to\mu\bar\mu$, $\tau\to\mu\mu\bar\mu$, $\mu\to e e \bar e$, $\tau\to \mu\gamma$, $\tau\to e\gamma$, $\mu\to e \gamma$, $b\to s\gamma$, $B\to D (D^*) \tau\nu$ and the rare higgs decay $h\to\tau\mu$ in order to determine the allowed parameter space of THDM-III and from these we give a prediction for the magnetic and weak magnetic dipole moment of charged  leptons, in particular of the $\tau$ lepton. The obtained  magnetic dipole moment of the $\tau$ lepton is of the order of $\mathcal{O}(10^{-8}-10^{-7})$ and the weak magnetic is of the order $\mathcal{O}(10^{-10}-10^{-7})$.

\end{abstract}

\pacs{Valid PACS appear here}
\maketitle

\section{\label{sec:level1}Introduction}

The study of dipole moments in particle physics has become an important tool both theoretically and experimentally in the search of new physics. Motivated in part by the fact that within the SM the electric dipole moment at the one loop level is zero, albeit different Beyond the Standard Model (BSM) theories predict a small yet different than zero value, this would serve as a new source of CP violation. In regard to the magnetic dipole moments, the theoretical prediction of the $a_{\mu}$ has been found to differ from the experimental value by \cite{MomMagMuonEXP}

\begin{eqnarray}
\Delta a_\mu = a_\mu^{Exp} - a_\mu^{SM} = 288(63)(49)\times 10^{-11},
\end{eqnarray}
\nolinebreak
which is greater than 3 standard deviations. This discrepancy between theory and experiment could give a hint to new physics. Of the several BSM models that have been used to try and offer a possible explanation to $\Delta a_\mu$, one of the simplest is the THDM,  as was done in the work of \cite{Pich} and \cite{Dumm}, which focused on  THDM's of the type I and type II. What differentiates each version of THDM is the way in which the Higgs doublets couple to fermions; in the type I only one of the doublets couples to all fermions while the other remains inert, type II has one doublet coupling exclusively to the up type quarks while the second doublet couples to down type quarks and  leptons. Finally, in the type III (THDM-III) both doublets couple to all fermions, up and down quarks and  leptons.  As far as the authors know, no one has determined if the THDM-III is a viable model to explain $\Delta a_\mu$ and in turn how it might constrain the parameter space of the model.
  
 Unlike the $\mu$, the $\tau$ lepton has a short life-time of $(290{.}3\pm 0{.}5)\times 10^{-15} s$ that does not allow measurement of its anomalous moments by the spin precession method used with the electron and muon. The dipole moment of the $\tau$, with a 95\% C.L., is \cite{delphi}
\begin{equation}
-0{.}052<a_{\tau}<0{.}013,
\end{equation}
with a central value 
\begin{equation}
a_{\tau}=-0{.}018(17),
\end{equation}
and the SM theoretical value is given by \cite{MomMagtauSM}
\begin{equation}
a_{\tau}^{SM}=1177{.}21(5)\times 10^{-6},
\end{equation}
\nolinebreak
different BSM models give predictions  for the magnetic moment of the $\tau$ between the range of $10^{-9}$ to $10^{-6}$ \cite{LQ},\cite{MSSM1},\cite{UP}. 

On the other hand, with respect to the weak properties of the $\tau$, the prediction of the SM was calculated in \cite{ValorSM-MDMDtau} whose value was found to be:
\begin{equation}
a_{\tau}^{W-SM}=-(2{.}10+0{.}61i)\times 10^{-6},
\end{equation}
\nolinebreak
different BSM models predict values between the range of $10^{-10}$ to $10^{-6}$ \cite{LQ},\cite{UP},\cite{MSSM3}. The real and imaginary part of the Weak Magnetic Dipole Moment (WMDM) of the $\tau$, with a 95\% C.L., were reported in \cite{MOMWEAK-EXP} and are given by   
\begin{equation}
Re\left(a_{\tau}^{W}\right) < 1{.}14\times10^{-3},\, Im\left(a_{\tau}^{W}\right) < 2{.}65\times10^{-3},
\end{equation} 
while the Weak Electric Dipole Moment (WEDM) is 
\begin{equation}
Re\left(d_{\tau}^{W}\right)  <  0{.}5\times10^{-17}ecm,\, Im\left(d_{\tau}^{W}\right)  < 1{.}1\times10^{-17}ecm.
\end{equation}
 The $\overline{f}fV$ effective vertex describing the interaction between a neutral vector boson and two on\textendash{}shell fermions can be  written as:
{\small{
\begin{equation}
\Gamma_{\mu}^{ffV}(q^{2})=ie\left[\gamma_{\mu}\left(F_{V}^{V}-F_{V}^{V}\gamma_{5}\right)-q_{\mu}\left(iF_{S}^{V}+F_{P}^{V}\gamma_{5}\right)+\sigma_{\mu\nu}q^{\nu}\left(iF_{M}^{V}+F_{E}^{V}\gamma_{5}\right)\right],
\end{equation}
}}where $q=p_{2}-p_{1}$ is the four-momentum.
The dipole moments are defined as follows:
\begin{equation}
a_{f}      = -2m_{f}F_{M}^{V=\gamma}(0),\,d_{f}      = -eF_{E}^{V=\gamma}(0),
\end{equation}
and the weak dipole moments 
\begin{equation}
a_{f}^{W}  =  -2m_{f}F_{M}^{V=Z}(m_{Z}^{2}),
d_{f}^{W}  =  -eF_{E}^{V=Z}(m_{Z}^{2}).
\end{equation}
  In this work, we focus on the contributions to magnetic and weak magnetic dipole moments of charged leptons that arise from a THDM-III with a four zero texture. In particular we study the anomalous magnetic dipole moment of the $\tau$  which can be measured through the decay $\tau^-\rightarrow\l^- \nu_{\tau} \overline{\nu_l}\gamma$  \cite{Tavares12} and its weak magnetic dipole moment \cite{chinos}.

 The structure of the paper is as follows; in section \ref{sec:THDM} we introduce the Lagrangian for the Yukawa sector of the THDM-III and how it can be expressed within the four zero texture framework. Afterwards in section \ref{sec:Dipoles} we present our calculations for the electromagnetic and weak dipole moments along with there one loop Feynman diagrams. While in section \ref{sec:Low energy} we present the effective Hamiltonians and the current values on experimental bounds and measurements for the Low energy processes considered. Results and numerical analysis are given in section \ref{sec:Numerical}. Finally, our conclusions are presented in section \ref{sec:conclusion}.

\section{The Two Higgs Doublet Model with a Four Zero Texture}\label{sec:THDM}

In this section we obtain the coupling of the Higgs bosons to fermions  $H(H^{\pm})f_i \overline{f}_j$. We show the way in which we diagonalize the fermion mass matrix, where the structure of the Yukawa matrices  is such that after summing them we recover a mass matrix  with a 4 zero texture. 

The general THDM, has been studied under different theoretical frameworks, such as the Minimal Flavor violating (MFV) THDM \cite{Buras:2010mh}, which provides the minimal level of FCNC consistent with data. The THDM with Alignment is a scenario in which flavor violation does not occur at tree level; the contributions coming from vertices where flavor change occurs are not present and therefore could be smaller then the ones present in our model.

The assumption of textures implies a specific pattern of FCNC Higgs-fermion couplings, known as the Cheng-Sher ansatz $\frac{\sqrt{m_i m_j}}{v}$. These vertices satisfy FCNC bounds for Higgs masses lighter than O(TeV),  which is one of the reasons why we work in this formalism. Secondly, it gives the structure of the Yukawa matrices as a function of fermion masses in such a way that the elements of the CKM matrix are reproduced. But there are different possibilities for the texture forms, such as a textured six, five or four zero mass matrix which are  ideas that has been extensively developed; one of the first efforts to work in this scheme assumed a six zero texture type \cite{Cheng:1987rs}, nonetheless this texture has been ruled out\cite{Honda:2004qh}. In this work we focus on the four zero type motivated by the fact that they currently satisfy many phenomenological observations. Furthermore, for certain four zero texture types as the Nearest Neighbor Interaction (NNI) is obtained through a $Z_4$ flavor symmetry in \cite{Branco:2010tx}. For a more in depth study of the formalism of textures and the THDM, we suggest Refs. \cite{BrancoReview}, \cite{HNishiura}, \cite{papaqui}.

The Yukawa Lagrangian of the THDM-III given by:

\begin{eqnarray} \label{eq:Lagrangian}
 \mathcal{L}_Y^f&=& Y_1^u\overline{Q}_{L} \tilde{\Phi}_1 u_{R}  + Y_2^u \overline{Q}_{L} \tilde{\Phi}_2 u_{R} \\ \nn
 &+& Y_1^d \overline{Q}_{L} \Phi_1 d_{R} + Y_2^d \overline{Q}_{L} \Phi_2 d_{R}  \\ \nn
 &+& Y_1^l \overline{L}_{L} \Phi_1 l_{R} + Y_2^l \overline{L}_{L} \Phi_2 l_{R} + h.c.,
\end{eqnarray}

\noindent here $\Phi_i=(\phi_i^+,\,\phi_i^0)^T$ denote the Higgs doublets, $Y_i^f$ ($f=u,d,l$) are 3$\times$3 Yukawa matrices, $\tilde{\Phi}_j=i\sigma_2\Phi_j^*$ and $\overline{Q}_L=(\bar{u}_L,\,\bar{d}_L)$.
 Since both Higgs doublets couple to all fermions, the mass matrix for each fermion type $f$  receives contributions from both vevs $v_1$ and $v _2$, i.e.

\begin{equation}
{M_f}=\frac{1}{\sqrt{2}} \left(v_1 {Y}^f_1 + v_2 {Y}^f_2 \right).
\end{equation}
In order to obtain physical fermion masses, we need to diagonalize the mass matrix; this is achieved through a bi-unitary transformation which is given by:

\begin{equation}
{M_D}=\mathcal{U}_f{M_f}\mathcal{U}_f^\dagger = \mathcal{U}_f \frac{1}{\sqrt{2}} \left(v_1 {Y}^f_1 + v_2 {Y}^f_2 \right)\mathcal{U}_f^\dagger,
\end{equation}
\nolinebreak
where the diagonalization matrix that we are using is obtained from \cite{Branco:1999nb}

\begin{eqnarray}
\mathcal{U}_f = \mathcal{O}_f^\dagger P_{f},
\end{eqnarray}
\nolinebreak
in which $\mathcal{O}_f$ is of the form

\begin{eqnarray}
\small \mathcal{O}_f =
\left( \begin{array}{ccc}
\sqrt{\frac{m_2 m_3 (A -m_1 )}{A  (m_2-m_1 ) (m_3-m_1 )}}
& \sqrt{\frac{m_1 m_3 (m_2-A )}{A (m2-m_1 ) (m_3-m_2 )}}
& \sqrt{\frac{m_1m_2 (A-m_3 )}{A (m_3-m_1 ) (m_3-m_2 )}} \\
-\sqrt{\frac{m_1 (m_1-A )}{ (m_2-m_1 ) (m_3-m_1 )}}
& \sqrt{\frac{m_2 (A-m_2 )}{ (m_2-m_1 ) (m_3-m_2 )}}
& \sqrt{\frac{m_3 (m_3-A )}{ (m_3-m_1 ) (m_3-m_2 )}} \\
\sqrt{\frac{m_1 (A-m_2 ) (A-m_3 )}{A (m_2-m_1 ) (m_3-m_1 )}}
& -\sqrt{\frac{m_2 (A-m_1 ) (m_3-A )}{A (m_2-m_1 ) (m_3-m_2 )}}
& \sqrt{\frac{m_3(A-m_1 ) (A-m_2 )}{A (m_3-m_1 ) (m_3-m_2 )}}
\end{array} \right),
\end{eqnarray} \\

\noindent such that  $A=m_3(1-\gamma_f r_2)$ with $f=l,\,u,\,d$ and $r_2=\frac{m_2}{m_3}$, and

\begin{eqnarray}\label{matrizP}
P =
\left( \begin{array}{ccc}
1 & 0               & 0\\
0 & e^{i\theta_1} & 0\\
0 & 0               & e^{i\theta_2} \\
\end{array} \right).
\end{eqnarray}

The neutral Yukawa Lagrangian given in Eq. (\ref{eq:Lagrangian}) can be expressed in terms of the mass eigenstates $h^0$, $H^0$, $A^0$ as \cite{GomezBock:2005hc}

{\footnotesize{}
\begin{eqnarray}\label{lagrangiano}
\mathcal{L}_{Y}^{neu} & = & \frac{g}{2}\left(\frac{m_{l}}{m_{W}}\right)\bar{l}\left[\frac{\cos\alpha}{\cos\beta}\delta_{ll{'}}+\frac{\sqrt{2}\sin(\alpha-\beta)}{g\cos\beta}\left(\frac{m_{W}}{m_{l}}\right)\left(\tilde{Y}_{2}^{l}\right)_{ll{'}}\right]l{'}H^{0}\nonumber \\
 & + & \frac{g}{2}\left(\frac{m_{l}}{m_{W}}\right)\bar{l}\left[-\frac{\sin\alpha}{\cos\beta}\delta_{ll{'}}+\frac{\sqrt{2}\cos(\alpha-\beta)}{g\cos\beta}\left(\frac{m_{W}}{m_{l}}\right)\left(\tilde{Y}_{2}^{l}\right)_{ll{'}}\right]l{'}h^{0}\\
 & + & \frac{ig}{2}\left(\frac{m_{l}}{m_{W}}\right)\bar{l}\left[-\tan\beta\delta_{ll{'}}+\frac{\sqrt{2}}{g\cos\beta}\left(\frac{m_{W}}{m_{l}}\right)\left(\tilde{Y}_{2}^{l}\right)_{ll{'}}\right]\gamma^{5}l{'}A^{0}\nonumber \\
 & + & \frac{g}{2}\left(\frac{m_{u}}{m_{W}}\right)\bar{u}\left[\frac{\sin\alpha}{\sin\beta}\delta_{uu{'}}+\frac{\sqrt{2}\sin(\alpha-\beta)}{g\sin\beta}\left(\frac{m_{W}}{m_{u}}\right)\left(\tilde{Y}_{2}^{l}\right)_{uu{'}}\right]u{'}H^{0}\nonumber 
\end{eqnarray}
 }

{\footnotesize{}
\begin{eqnarray*}
 & + & \frac{g}{2}\left(\frac{m_{u}}{m_{W}}\right)\bar{u}\left[-\frac{\cos\alpha}{\sin\beta}\delta_{uu{'}}+\frac{\sqrt{2}\cos(\alpha-\beta)}{g\sin\beta}\left(\frac{m_{W}}{m_{u}}\right)\left(\tilde{Y}_{2}^{u}\right)_{uu{'}}\right]u{'}h^{0}\\
 & + & \frac{ig}{2}\left(\frac{m_{u}}{m_{W}}\right)\bar{u}\left[-\cot\beta\delta_{uu{'}}+\frac{\sqrt{2}}{g\sin\beta}\left(\frac{m_{W}}{m_{u}}\right)\left(\tilde{Y}_{2}^{u}\right)_{uu{'}}\right]\gamma^{5}u{'}A^{0},
\end{eqnarray*}
}

\noindent where $\tilde Y_2^f=\frac{\sqrt{2}}{v_2}M_D-\frac{v_1}{v_2}\tilde Y_1^f$ and $ \tilde Y_i^f= \mathcal{U}_f  {Y}^f_i\mathcal{U}_f^\dagger$. The type-down quarks part is similar to the lepton part with the exchange $l\to d$ and $m_l\to m_d$.

On the other hand, the charged Yukawa Lagrangian is given by:
\begin{eqnarray}\label{lagrangianoCH}
\nonumber \mathcal{L}_{Y}^{ch} & = & \left[\bar{d}_{i}\left(\tilde{Y}_{1}^{u}sin\beta+\tilde{Y}_{2}^{u}cos\beta\right)u_{j}H^{-}\right.\\ \nonumber
 & + & \left.\bar{u}_{i}\left(\tilde{Y}_{2}^{d}cos\beta-\tilde{Y}_{1}^{d}sin\beta\right)d_{j}H^{+}\right]P_{R}\\
 & + & \left[\bar{d}_{i}\left(\tilde{Y}_{2}^{u}cos\beta-\tilde{Y}_{1}^{u}sin\beta\right)u_{j}H^{-}\right.\\ \nonumber
 & + & \left.\bar{u}_{i}\left(\tilde{Y}_{1}^{d}sin\beta-\tilde{Y}_{2}^{d}cos\beta\right)d_{j}H^{+}\right]P_{L}. 
\end{eqnarray}
We will use a notation in which $\eta_{f\bar{f}}^H$ corresponds to the couplings $f\bar{f}H$ in the Lagrangians, for example:
\begin{equation}
\eta_{l\bar{l}}^{H^0}=\frac{g}{2}\left( \frac{m_l}{m_W} \right)  \left[\frac{\cos\alpha}{\cos\beta} \delta_{ll{'}} + \frac{\sqrt{2}\sin(\alpha-\beta)}{g\cos\beta} \left( \frac{m_W}{m_l} \right) \left( \tilde Y_2^l \right)_{ll{'}}\right].
\end{equation}

In this work, we assume that the mass matrix has a four zero texture form

\begin{eqnarray}
{M_f} =  \left( \begin{array}{ccc}
0 & D_f & 0 \\
D_f^* & C_f & B_f \\
0 & B_f^* & A_f
\end{array} \right),
\label{eq:mass}
\end{eqnarray}

\noindent so

\begin{eqnarray}
 \left( \begin{array}{ccc}
0 & D_f & 0 \\
D_f^* & C_f & B_f \\
0 & B_f^* & A
\end{array} \right) = \frac{v_1}{\sqrt{2}}\left( \begin{array}{ccc}
0 & d_1 & 0\\
d^*_1 & c_1 & b_1\\
0 & b^*_1 & a_1 \\
\end{array} \right) + \frac{v_2}{\sqrt{2}}\left( \begin{array}{ccc}
0 & d_2 & 0\\
d^*_2 & c_2 & b_2\\
0 & b^*_2 & a_2 \\
\end{array} \right),
\label{eq:mass}
\end{eqnarray} \\
where

\[
Y_{i}=\left(\begin{array}{ccc}
0 & d_{i} & 0\\
d_{i}^{*} & c_{i} & b_{i}\\
0 & b_{i}^{*} & a_{i}
\end{array}\right),\: i=1,\,2.
\]

To find the expression for the mass eigenstates, we use the following matrix invariants:

\begin{eqnarray}
Tr(M_f) &=& C_f+A_f=m_{f_{3}}+m_{f_{2}}+m_{f_{1}},  \\
det(M_f) &=& -D_f^2 A_f = m_{f_{1}} m_{f_{2}} m_{f_{3}},  \\
\chi(M_f) &=& B_f^2+C_fA_f - D_f^2  = m_{f_{1}}m_{f_{2}} + m_{f_{1}}m_{f_{3}} + m_{f_{2}}m_{f_{3}},
\end{eqnarray}
\nolinebreak
from these expressions we find a relation between the components of the 4-texture mass matrix and the physical fermion masses

\begin{eqnarray}
A_f &=& m_{f_{3}} \left(1-r_2 \gamma_f \right), \\
B_f &=& m_{f_{3}} \sqrt{\frac{r_2 \gamma_f  \left(r_2 \gamma_f +r_1-1\right) \left(r_2 \gamma_f +r_2 - 1\right)}{1-r_2 \gamma_f}}, \\
C_f &=& m_{f_{3}} \left(r_2 \gamma_f +r_1+r_2\right), \\
D_f &=& \sqrt{-\frac{m_{f_{1}} m_{f_{2}}}{1-{r_2} \gamma_f }}, 
\end{eqnarray}
where $m_{f_{i}}$ is the fermion mass and the $r_i$ stands for $r_i=\frac{m_{f_{i}}}{m_{f_{3}}}$.

So far, we have found a relation between the textured mass matrix and the diagonal physical mass matrix, but the coupling of fermions to the scalar is given through the Yukawas, whose elements $a_{1,2}$, $b_{1,2}$, $c_{1,2}$ and $d_{1,2}$ can not be  simultaneously determined. But instead of having them as free parameters we can do the following:

Case 1) We will assume that the elements of $Y_1$ are negligible and so

\begin{eqnarray}
 M_f =  \left ( \begin{array}{ccc}
0 & D_f & 0 \\
D_f^* & C_f & B_f \\
0 & B_f^* & A
\end{array} \right) \approx  \frac{v_2}{\sqrt{2}}\left( \begin{array}{ccc}
0 & d_2 & 0\\
d^*_2 & c_2 & b_2\\
0 & b^*_2 & a_2 \\
\end{array} \right).
\label{eq:mass}
\end{eqnarray} \\

Case 2a) Following \cite{MarcoKikeTHDMTX}, we will consider the case in which each Yukawa contributes to only one element of the mass matrix

\begin{eqnarray}
 M_f =  \left ( \begin{array}{ccc}
0 & D_f & 0 \\
D_f^* & C_f & B_f \\
0 & B_f^* & A
\end{array} \right) = \frac{v_1}{\sqrt{2}}\left( \begin{array}{ccc}
0 & d_1 & 0\\
d_1^* & c_1 & b_1\\
0 & b_1^* & 0 \\
\end{array} \right) + \frac{v_2}{\sqrt{2}}\left( \begin{array}{ccc}
0 & 0 & 0\\
0 & 0 & 0\\
0 & 0 & a_2 \\
\end{array} \right),
\label{eq:mass}
\end{eqnarray} \\

or

Case 2b)
\begin{eqnarray}
 M_f =  \left ( \begin{array}{ccc}
0 & D_f & 0 \\
D_f^* & C_f & B_f \\
0 & B_f^* & A
\end{array} \right) = \frac{v_1}{\sqrt{2}}\left( \begin{array}{ccc}
0 & 0 & 0\\
0 & c_1 & 0\\
0 & 0  & 0 \\
\end{array} \right) + \frac{v_2}{\sqrt{2}}\left( \begin{array}{ccc}
0 & d_2 & 0\\
d_2^* & 0 & b_2\\
0 & b_2^* & a_2 \\
\end{array} \right).
\label{eq:mass}
\end{eqnarray}

in both cases, the structure of the Yukawa matrices textures is such that after summing both we recover a mass matrix with a 4 zero texture. In working under any of these assumptions we avoid introducing new free parameters instead of explicitly determining the Yukawa matrix elements as functions of the fermion masses and $\gamma_f$. The $\left(\tilde{Y}_{2}\right)_{ff'}$ for the paralell case is shown in the App.\ref{app:chis-cases1-6} and for cases 2a and 2b are given in \cite{MarcoKikeTHDMTX}.

In regard to the Renormalization Group Equations (RGE) stability, a study of the RGE in the general THDM-III has already been performed by the authors of \cite{RGETHDM}, they found a remarkable stability of the suppressed FCNC Yukawa coupling parameters. Furthermore, in \cite{RGETHDM4zeros} they explored if four zero textures are stable under the running of energy scales and found that
the texture zeros of mass matrices $M_f$ are stable against the evolution of energy scales.


 \section{DIPOLE MOMENTS IN THE THDM-III with four zero texture}\label{sec:Dipoles}

In this section we give our expressions for the magnetic and weak magnetic dipole moments of charged leptons within the THDM-III.

\subsection{MUON ANOMALOUS MAGNETIC DIPOLE MOMENT}

\begin{figure}[H]
\centering
    \subfigure[ ]{\includegraphics[width = 0.35\textwidth]{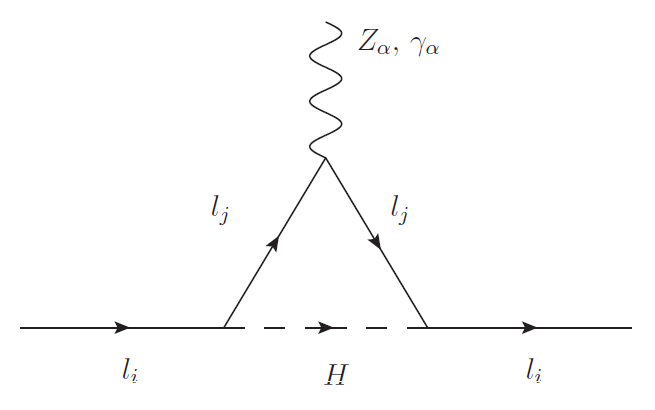}}\label{fig:loop1}
     \subfigure[ ]{\includegraphics[width = 0.35\textwidth]{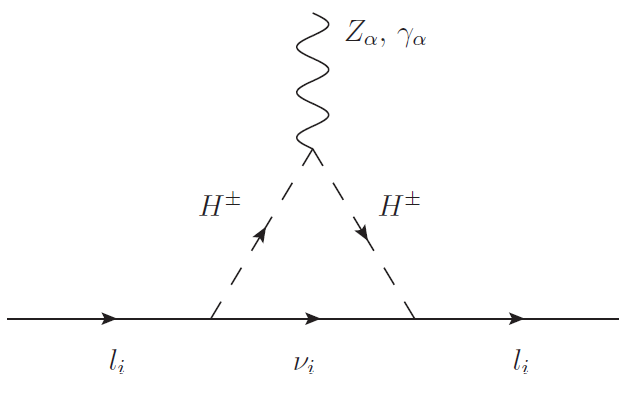}}\label{b}
     \subfigure[ ]{\includegraphics[scale=0.36]{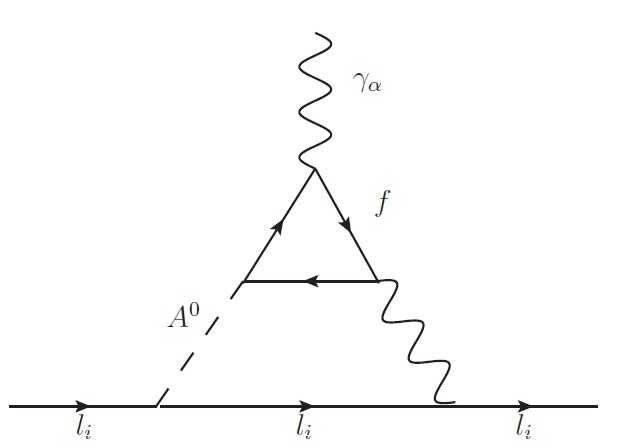}}\label{Barr}   
      \caption{Feynman diagrams that contribute to the magnetic dipole moment of charged leptons in the framework of the THDM-III. In the case of the contributions coming of Barr-Zee diagrams we only consider the dominant contribution given by diagram (c).  \label{fig:loop-a-b}}
	\end{figure}

The Feynman diagrams that contribute to the magnetic dipole moment of charged leptons in the framework of the THDM-III are shown in FIG.[\ref{fig:loop-a-b}],

\noindent where $l_i=e$, $\mu$, $\tau$, and $l_j$ refers to a lepton of different flavor, $H=h^0, H^0,A^0$ are the neutral light, heavy and pseudo scalar Higgs bosons, respectively. The vertices coupling Higgs bosons to charged fermions are obtained from the Lagrangian in eqs.(\ref{lagrangiano}-\ref{lagrangianoCH}), except for the vertices coupling Higgs bosons to  gauge bosons and Z to neutrinos \cite{Gunion:1989we}. The contributions to the magnetic dipole moments of leptons are CP-conserving, therefore the phases $\theta_{1,\,2}$ are set to zero.

We perform the calculation in the unitary gauge via the Feynman parameters. The one-loop contribution is given by:

\begin{equation}
a_{l_{i}}^{\rho_{1}\rho_{2}\rho_{3}}=\underset{\underset{l'=2}{l=1,\,2,\, 3}}{\sum}\frac{\left|\eta_{_{ll'}}^{\rho_{1}\rho_{2}\rho_{3}}\right|^2 m_{l_i}}{\sqrt{8}\pi^{2}}\intop_{0}^{1}dx\intop_{0}^{1-x}dy\,G_{k}^{\rho_{1}\rho_{2}\rho_{3}}(x,\, y),
\end{equation}
\nolinebreak
where the label $\rho_{n}$ refers to the particles circulating in each diagram, and the index $ll'$ denotes the entry of the Yukawa interaction, in the case for the $a_{\mu}$: $l'=2$, i.e. $l2$ with $l=1,\,2,\,3$.  Notice that our results can be applied to any of the charged lepton. The label $\eta_{_{ll'}}^{\rho_{1}\rho_{2}\rho_{3}}$ is the coupling obtained from the neutral Lagrangian term of $l\bar{l}H$. \\

The $G_{k}^{\rho_{1}\rho_{2}\rho_{3}}(x,\, y)$ function  for the diagram (a) is given by:

\begin{equation}\label{eq1}
G_{a}^{{H} l_{j} l_{j}}(x,\,y)=(x+y)(m_{l_{j}}-m_{l_{i}}(x+y-1))/M_{a}^{2},
\end{equation}
\nolinebreak
with $M_{a}^{2}=-m_{H}^{2}(x+y-1)+(x+y)(m_{l_{j}}^{2}+m_{l_{i}}^{2}(x+y-1)$. Here $l_{j}=e,\,\mu,\,\tau.$
 For the Feynman diagram (b): $\eta_{_{ll'}}^{\rho_{1}\rho_{2}\rho_{3}}=1$  and $G_{b}^{\rho_{1}\rho_{2}\rho_{3}}(x)$ function is given by:
\begin{equation}\label{eq2}
G_{b}^{H^{\pm}H^{\pm}\nu}(x)=2m_{l_{i}}x/M_{b}^{2},
\end{equation}

\noindent with $M_{b}^{2}=(m_{l_{i}}^{2}x-m_{H^{\pm}}^{2})$.
The THDM also has Barr-Zee two-loop contributions to the muon anomalous magnetic moment. The dominant contribution is given by \cite{Barr-Zee}
\begin{equation}
a_{\mu}^{two\,loop}=\frac{\alpha^2}{8\pi^2s_W^2}\frac{m_{\mu}^2 \eta_{\mu\bar{\mu}}^{A^0}}{m_W^2} \sum_{f=t,\,\tau,\,b} N_c^fQ_f^2r_ff(r_f)\eta_{f\bar{f}}^{A^0},
\end{equation}
where $r_f=(m_f/m_A^0)$, $m_f$ is the fermion mass, $N_c^f=1(3)$ for leptons (quarks), $Q_f$ is the electric charge of fermions and $\eta_{f\bar{f}}^{A^0}$ is given by the Lagrangian coupling the CP-odd scalar to fermions:
\begin{equation}
\mathcal{L}=i\frac{g\,m_f\,\eta_{f\bar{f}}^{A^0}}{2m_W}\bar{f}\gamma^5fA^0,
\end{equation}
and finally 
\begin{equation}
f(x)=\int_0^1 \frac{log(\frac{x}{y(1-y)})}{x-y(1-x)}dy.
\end{equation}

Because of the term $r_f=(m_f/m_A^0)$ there is an inverse mass dependence with the pseudo-scalar, which in our work was evaluated in the range of $200-1000\,GeV$, hence the contribution of this diagram is of the order $\mathcal{O}(10^{-13})$. In a previous study by \cite{TavCP},  the Barr-Zee contributions are considerable because they consider pseudo-scalar masses between $10^{-2}-10^2\,GeV$. However, a collider search for the neutral Higgs bosons in the framework of the THDM's rules out little masses for $m_{A^0}$. An OPAL analysis \cite{OPAL}  using the LEP-II 
data up to $\sqrt{s}=189GeV$ exclude the region $12<m_A<56\, GeV$.

Therefore, the total contribution to $a_{\mu}$ in the THDM-III is

\begin{equation}
a_{\mu}^{THDM-III}=a_{\mu}^{H^{\pm}H^{\pm}\nu}+\underset{\underset{l_{j}=e,\,\mu,\,\tau}{H=h^0,\, H^0,\, A^0}}{\sum}a_{\mu}^{H l_{j}l_{j}}+a_{\mu}^{two\,loop}.
\end{equation}

\subsection{Magnetic and weak magnetic dipole moments of the $\tau$}\label{sec:WeakTau}

The analysis that follows can also be applied to any of the charged leptons $l_i=e$, $\mu$, $\tau$. Nonetheless, we will concentrate on the case $l_i=\tau$. The Feynman diagrams that contribute to the magnetic dipole moment of the $\tau$ are similar to the ones used for the muon, except that we now have that $m_{l_i}=m_{\tau}$ and $l'=3$ because the entries of $\eta_{_{ll'}}^{\rho_{1}\rho_{2}\rho_{3}}$ are $l3$, i.e. $13,\,23,\,33$.

\begin{figure}[h]
\centering
\addtocounter{subfigure}{3}
\subfigure[ ] {\includegraphics[width = 0.3\textwidth]{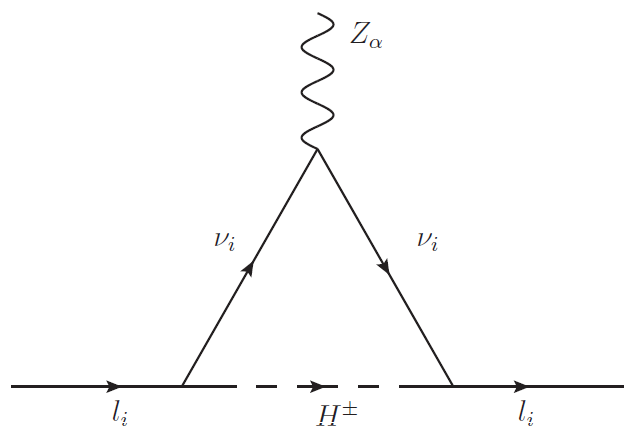}}
\subfigure[ ] {\includegraphics[width = 0.3\textwidth]{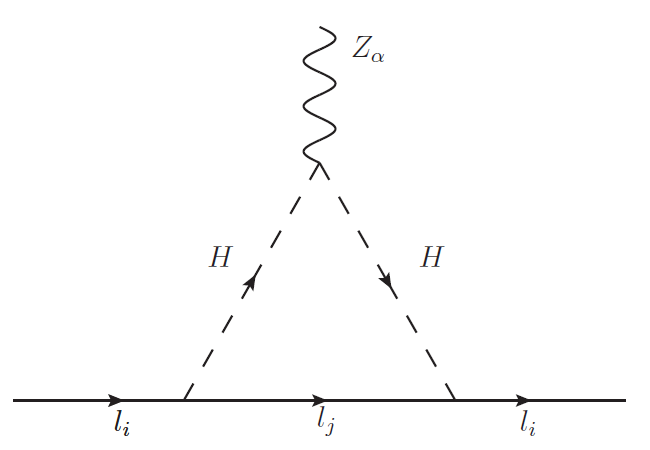}}
\subfigure[ ] {\includegraphics[width = 0.3\textwidth]{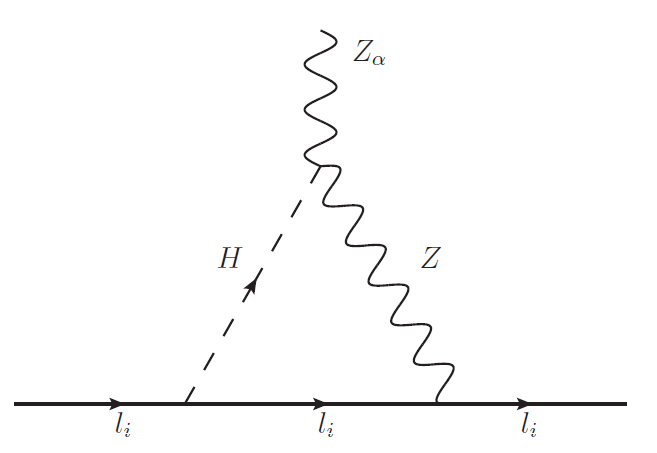}}
\caption{Additionally to the Feynman diagrams (a) y (b) there are contributions from (d), (e) y (f) to the weak dipole moment of the charged leptons $a^W_{l_i}$. In the Feynman diagrams (e) $H=h^0,\,H^0,\,A^0$ and (f) $H=h^0,\,H^0.$ \label{debiles}}
	\end{figure}

For the weak magnetic dipole moment, we also use diagrams like those shown in FIG.[\ref{fig:loop-a-b}] with the photon replaced by the Z boson, but we must also consider the diagrams shown in FIG.[\ref{debiles}]

The contributions to WMDM of the charged leptons of each Feynman diagram  are given by:

\begin{equation}
\left(a_{l_i}^{W}\right)^{\rho_{1}\rho_{2}\rho_{3}}=\sum_{a,\, b,\,c,\, e}A_{a,\, b,\,c,\, e}\intop_{0}^{1}dx\intop_{0}^{1-x}dy\, G_{k}^{\rho_{1}\rho_{2}\rho_{3}}(x,\, y),
\end{equation}
\nolinebreak
again the label $\rho_{n}$ refers to the particles circulating in a particular one-loop diagram, $A_{a,\, b,\,\,c,\, e}$ represent terms associated to each Feynman diagram, given by

\begin{eqnarray}
A_{a} & = & \frac{m_{\tau}\left|\eta_{ll'_{a}}^{\rho_{1}\rho_{2}\rho_{3}}\right|^{2}}{4\pi^{2}c_{W}s_{W}},\\
A_{b} & = & \frac{m_{\tau}^{3}g_{W}^{3}(c_{W}^{2}-s_{W}^{2})}{32\pi^{2}m_{W}^{2}c_{W}},\\
A_{d} & = & \frac{g_{W}^{2}g_{Z}}{128\pi^{2}m_{W}^{2}c_{W}},\\
A_{f} & = & \frac{m_{\tau}em_{Z}\eta_{lm_{f}}^{\rho_{1}\rho_{2}\rho_{3}}}{4\pi^{2}c_{W}^{3}s_{W}^{2}}cos(\alpha-\beta), 
\end{eqnarray}
\nolinebreak
 where $c_W,\,s_W,\,m_W,\,m_Z$ are the cosine, sine of the Weinberg angle, W and Z gauge boson mass, respectively.
 We omit $A_{e}$ because the Feynman diagram (e) gives no contribution to weak magnetic dipole moment.
 The $G_{k}^{\rho_{1}\rho_{2}\rho_{3}}(x,\, y)$
functions are:

\begin{eqnarray}
G_{a}^{Hl_{j}l_{j}}(x,\, y) & = & -2(C_{V}(x+y)(m_{l_{i}}(x+y-1)-m_{l_{j}}))/M_{a}^{2},\\
G_{b}^{H^{\pm}H^{\pm}\nu}(x,\, y) & = & 4m_{l_{i}}x(x-1)/M_{b}^{2}, \\
G_{d}^{H^{\pm}\nu\nu}(x,\, y) & = & -8m_{l_{i}}(x+y-1)(x+y)/M_{d}^{2}, \\
G_{f}^{ZH\tau}(x,\, y) & = & -\frac{C_{V}}{m_{Z}^{2}M_{f}^{2}}(-m_{l_{i}}^{2}((x+y-1)^{2}(x+y-1)\\
 & + & M_{f}^{2}(\log(M_{f}^{2})(2-3(x+y))-x-y-1)+m_{Z}^{2}x(y(x+y-1)+2)),\nonumber
\end{eqnarray}

where
\begin{eqnarray}
M_{a}^{2} & = & m_{l_{i}}^{2}(x+y-1)(x+y)+m_{l_{j}}(x+y)-m_{Z}^{2}xy-m_{H}^{2}(x+y-1),\\
M_{b}^{2} & = & m_{l_{i}}^{2}(x-1)x+m_{Z}^{2}y(x+y-1)-m_{H}^{2}(x-1),\\
M_{d}^{2} & = & m_{l_{i}}^{2}(x+y-1)(x+y)-m_{Z}^{2}xy-m_{H}^{2}(x+y-1),\\
M_{f}^{2} & = & m_{l_{i}}^{2}((x+y-1)(x+y-1))-m_{Z}^{2}x(y-1)+m_{H}^{2}y.
\end{eqnarray}
The total contribution to $(a_\tau^W)$ is:
\begin{equation}
(a_\tau^W)^{THDM-III}=(a_{\tau}^W)^{H^{\pm}H^{\pm}\nu}+\underset{\underset{l_{j}=e,\,\mu,\,\tau}{H=h^0,\, H^0,\, A^0}}{\sum}(a_{\tau}^W)^{H l_{j}l_{j}}+(a_{\tau}^W)^{H^{\pm}\nu\nu}+(a_{\tau}^W)^{ZH\tau}.
\end{equation}

\section{Constraints from Low Energy Processes}\label{sec:Low energy}
Due to the introduction of a second scalar doublet, there are additional Feynman diagrams that are not present in the SM, involving the exchange of  Higgs bosons ($h^0,\,H^0,\,A^0,\,H^{\pm}$). There are currently strict bounds and measurements that have been obtained through various low energy phenomenological studies, this allows us to constraint parameters of the model such as $\tan{\beta}(=t_{\beta})=v_2/v_1$ which controls the intensity of the coupling of the Higgs bosons to the fermions. In this section we show the allowed parameters space in the plane $t_{\beta}-m_{H(H^{\pm})}$, the other parameters are fixed. In particular for $\alpha$ and $\beta$ angles we study the scenario where  $\alpha-\beta = \frac{\pi}{2}$, which reproduces the case where the coupling of $h^0$ to fermions is SM-like. In general, the favored region for THDM's are phenomenologically viable for the scenario $\alpha-\beta = \frac{\pi}{2}$ \cite{alpha-beta}. We consider the relevant low energy constraints and the muon anomalous magnetic dipole moment for constraint the model parameters and then we give analytical predictions to magnetic and weak magnetic dipole moments of charged leptons in the section[\ref{sec:Numerical}] using  values of the allowed region. For the low energy constraints mediated by neutral Higgs bosons, we consider  $B_s^0 \to \mu\bar{\mu}$, $\mu\rightarrow e \gamma$, $\tau \rightarrow \mu \gamma$, $\tau \rightarrow e \gamma$, $K-\bar{K}$ mixing  and $l_i\rightarrow l_j l_k \overline{l_k}$ and the rare Higgs decay $h\to\tau\mu$. While for processes mediated by charged Higgs bosons $b\to s\gamma$ and $B\to D(D^*)\tau\nu$.
The processes $H^0\to 2\,jets,\,Z\to f\bar{f},\,e^+e^-\to H^0Z,\,H^0A^0,\,e^+e^-\to b\bar{b}H$  impose a limit on the mass of the Higgs boson given by $m_{H^0}>110.6\,GeV$ at 95\% \cite{mHcota} and in the range $1-55\,GeV$ is excluded \cite{mHcotainferior}.

On the other hand, a theoretical approach that can be used to try and understand the underlying physics is through the effective Hamiltonian, 
\begin{equation}
\mathcal{H}_{eff} = \frac{G_F}{\sqrt{2}}\sum_i V_{CKM}^i C_i(\mu)Q_i,
\end{equation}
where $Q_i$ are the operators which govern the decay and $C_i(\mu)$ are Wilson Coefficients. The scale $\mu$ separates the physics contributions into short distance contributions contained in $C_i(\mu)$ while the long distance contributions are contained in $Q_i$ but the Wilson Coefficients depend on the couplings of the particular model, which in our case are functions of $\alpha$, $\beta$, $\theta_i$, $m_{H(H^{\pm})}$ and $\gamma_f$. This will allow us to constrain the model parameters.


\subsection{Low energy constraints mediated by neutral Higgs bosons}
\subsubsection{ $B_s^0\to\mu\bar\mu$}\label{sec:Bto2mu}

\begin{figure}[H]
\centering
\includegraphics[scale = 0.4]{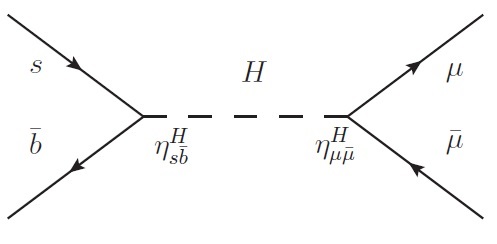}
\caption{ Feynman diagram for the $B_s^0\to\mu\bar{\mu}$ process at the quark level. The labels $\eta_{f\bar{f}}^H$  indicate the
new contributions coming from the THDM-III. We will omit these labels in all further Feynman diagrams.
\label{fd:Btomumu}}
\end{figure}

 A measurement of $B_s^0\rightarrow\mu\overline{\mu}$  was reported in \cite{Khachatryan}, whose value is
 \begin{equation}
 BR(B_s^0\rightarrow\mu\bar{\mu})=(2{.}8_{-0{.}6}^{+0{.}7})\times10^{-9},
\end{equation}
while the SM value is given by \cite{BmumuSM}
\begin{equation}
 BR(B_s^0\rightarrow\mu\bar{\mu})_{SM}=(3{.}23\pm0{.}27)\times10^{-9},
\end{equation}
\nolinebreak
 which puts very stringent constraints on the model.  The Feynman diagram for this process at the quark level is shown in FIG.[\ref{fd:Btomumu}].

We observe that the new contributions from Higgs bosons appear in the $H\bar{s}b$ and $H\mu\bar{\mu}$ couplings.
The Branching ratio for this decay is given by \cite{Dedes},

	\begin{eqnarray}
\nn &BR&\left(B_{s}^0\rightarrow\bar{\mu}\mu\right) =
\frac{G_{F}^{2}\alpha_{em}^{2}}{16\pi^{3}}M_{B}\tau_{B}\left|V_{tb}V_{ts}^{*}\right|^{2}\sqrt{1-\frac{4m_{\mu}^{2}}{M_{B}^{2}}} \\
 &\times& \left[\left|F_{H}\right|^{2}\left(1-\frac{4m_{\mu}^{2}}{M_{B}^{2}}\right)+\left| F_{A^0}\right|^{2}\right],
 	\end{eqnarray}

\noindent where $F_{H}$ and $F_{A^0}$ are the form factors given by,

	\begin{eqnarray}
 F_{H,A^0}&=&-\frac{i}{2}M_{B}^{2}\, f_{B_{s}}\frac{m_{b}}{m_{b}+m_{s}}C_{H,A^0}, \nn\\
 	\end{eqnarray}

\noindent where $ G_{F}$ is the Fermi constant, $\tau_{B}$ the lifetime of the $B$ meson, $M_B$ the B meson mass, $f_{B_{s}}=0{.}256$   and  $C_{H},\;\; C_{A^0}$
the Wilson coefficients  that  appear in the effective Hamiltonian,

\begin{equation}
H_{eff}=-2\sqrt{2}G_{F}V_{tb}V_{ts}^{*}\sum_i C_i Q_i,
\end{equation}

\noindent where $Q_i$ are the effective operators and are given by:

	\begin{equation} \nonumber
Q_{H}=\frac{e^{2}}{16\pi^{2}}m_{b}(\bar{s\,}P_{R}b)(\bar{\mu}\, \mu),\; Q_{A^0}=\frac{e^{2}}{16\pi^{2}}m_{b}(\bar{s\,}P_{R}b)(\bar{\mu}\gamma^{5}\mu)
	\end{equation}

\noindent and the Wilson coefficients $C_i$ are,

\begin{equation}\label{equYukawa1} 
C_{H}= \frac{2\pi m_{\mu}}{V_{ts}^{*}V_{tb}\alpha_{em}}\Bigg[\frac{1}{4m_{H^{0}}^{2}}\sum_{H=H^0,\,h^0}\eta_{\bar{b}s}^H \eta_{m\bar{m}}^H
\Bigg],\;C_{A^0}=\frac{2\pi m_{\mu}}{V_{ts}^{*}V_{tb}\alpha_{em}} \Bigg[\frac{1}{4m_{A^{0}}^{2}} \eta_{\bar{b}s}^{A^0} \eta_{m\bar{m}}^{A^0}
 \Bigg],   
	\end{equation}	
	\nolinebreak where $m_b,\,V^*_{tb(ts)}, m_{H^0}, m_{h^0}, m_{A^0}$ are
bottom quark mass, CKM elements, heavy, SM-like, pseudo scalar Higgs bosons mass,
respectively. We can see through \ref{equYukawa1} the dependence on the model parameters since $\eta_{f\bar{f}}^{H,A^0}=\eta_{f\bar{f}}^{H,A^0}(\alpha,\,\beta,\,\theta_i,\,\gamma_f)$ and therefore $C_{H,A^0}=C_{H,A^0}(\alpha,\,\beta,\,\theta_i,\,m_H,\gamma_f)$.

\subsubsection{Radiative decays $l_i\to l_j \gamma$}\label{sec:mutoegamma}
\begin{figure}[H]
\centering
\includegraphics[scale = 0.4]{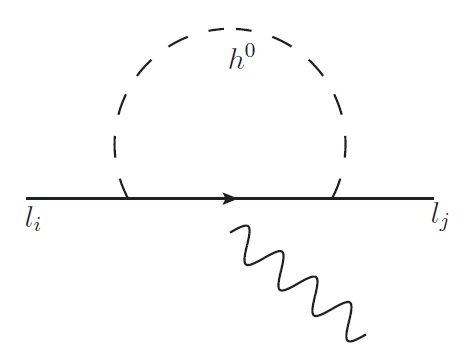}
\caption{ Feynman diagram for the $l_i \to l_j \gamma$ process.
\label{li-lj}}
\end{figure}

The second processes that we considered are the radiative decays $l_i\to l_j\gamma$. The Feynman diagram with the largest contributions for this process is shown in FIG.[\ref{li-lj}],

The branching ratio for the general decay of a lepton ($l_i$) to a lepton of a different family ($l_j$) is given by: \cite{crivellin}

\begin{eqnarray}
Br(l_i\to l_j\gamma) = \frac{m_{l_i}^5}{4\pi\Gamma_{l_i}}\left(\big|C_R^{l_jl_i}\big|^2 +
\big|C_L^{l_jl_i}\big|^2\right),
\end{eqnarray}
\nolinebreak
where the Wilson Coefficients $C_{R,L}^{l_jl_i}$ are:

\begin{eqnarray}
\nn C_{RH}^{l_jl_i} &=& \Bigg[
\eta_{l_il_j}^{LRH\star} \eta_{l_il_j}^{LRH} +
\eta_{l_jl_i}^{LRH\star} \eta_{l_jl_i}^{LRH} \\
&-& \frac{m_{l_i}}{m_{l_i}}\eta_{l_il_j}^{LRH}\eta_{l_jl_i}^{LRH}
\left(9 + 6\ln\left(\frac{m_{l_j}^2}{m_{H}^2}\right) \right) \Bigg],
\end{eqnarray}
\nolinebreak
such that $H=h^0, H^0, A^0$. $C_L^{l_jl_i}$ is obtained by simply interchanging R by L.

The experimental bounds for each of the different processes are \cite{36}, \cite{37}, \cite{38}
\begin{eqnarray*}
BR(\tau\rightarrow\mu\gamma) & \leq & 4{.}5\times10^{-8}\\
BR(\tau\rightarrow e\gamma) & \leq & 1{.}1\times10^{-7}\\
BR(\mu\rightarrow e\gamma) & \leq & 5{.}7\times10^{-13}
\end{eqnarray*}

Again we can see that $C_{RH}^{l_jl_i}=C_{RH}^{l_jl_i}(\alpha,\,\beta,\,\theta_i,\,m_H,\gamma_f)$.
\subsubsection{ $K-\bar{K}$ mixing}\label{sec:KKmixing}

\begin{figure}[H]
\centering
\includegraphics[scale = 0.4]{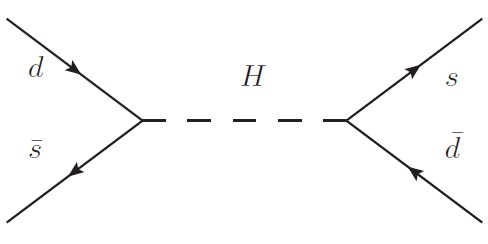}
\caption{ Feynman diagram for the $K\bar{K}-mixing$.
\label{fd:li-lj}}
\end{figure}

The amplitude for $K\bar{K}$ Mixing receives contributions from the Feynman diagram  shown in FIG.[\ref{fd:li-lj}].

The effective hamiltonian for this process at the quark level is written from \cite{Dedes:2002er} as follows

\be
H_{eff}^{\Delta S =2} = \frac{G_F^2 M_W^2}{16\pi^2}\sum_i C_i Q_i,
\ee

with $(i=L, R, LR)$, where

\begin{eqnarray}
Q_L^{\overline{s}d,s\overline{d}} &=& \big(\bar{s} P_L d\big) \big(\bar{s} P_L d\big),\\
Q_R^{\overline{s}d,s\overline{d}} &=& \big(\bar{s} P_R d\big) \big(\bar{s} P_R d\big),\\
Q_{LR}^{\overline{s}d,s\overline{d}} &=& \big(\bar{s} P_L d\big) \big(\bar{s} P_R d\big), 
\end{eqnarray}

and the Wilson coefficients are written here for a general multi-Higgs doublet model,
\be
C_L^{\overline{s}d,s\overline{d}} = - \frac{16 \pi^2}{G_F^2 m_W^2}m_sm_d |\eta_{sb}^H|^2 \sum_{a=1}^2 \frac{\big(U_{1a}^*\big)^2 }{m_{H}^2},\\
C_R^{\overline{s}d,s\overline{d}} = - \frac{16 \pi^2}{G_F^2 m_W^2}m_sm_d |\eta_{sb}^H|^2 \sum_{a=1}^2 \frac{\big(U_{1a}\big)^2 }{m_{H}^2},\\
C_{LR}^{\overline{s}d,s\overline{d}} = - \frac{16 \pi^2}{G_F^2 m_W^2}m_sm_d |\eta_{sb}^H|^2 \sum_{a=1}^2 \frac{U_{1a}^* U_{1a} }{m_{H}^2},
\ee

\noindent where $H=H^0,\,h^0,\,A^0$. $U_{ba}$ is the rotation matrix to go from weak to mass-eigenstates in the neutral CP-even Higgs sector,  since we are working in general THDM, the rotation matrix for neutral Higgs bosons takes the form

\be
U=
 	\begin{pmatrix}
\cos\alpha&-\sin\alpha\\
\sin\alpha&\cos\alpha\\
  	\end{pmatrix}.
\nn
\ee

Flavor violation arises from the non-diagonal terms of the K (neutral kaon) mass matrix, in particular the component
$M_{12}^K,$  whose experimental value has been measured to be

\be
M_{12}^K = \frac{\Delta M_K}{M_K} = 7{.}2948 \times 10^{-15} GeV,
\ee

\noindent but given that $\Delta M_K$ is obtained through

	\begin{eqnarray}
\nn \Delta M_K &=& 2 Re\langle \bar{K}^0|H_{eff}^{\Delta S = 2}|K^0\rangle \\ \nn
&=& \frac{G_F^2 M_W^2}{12 \pi^2} M_K F_K^2 \eta_2 {B}_K  \\ 
&\times&\left[ \bar{P}_{2,LR} C_{LR}^{\overline{s}d,s\overline{d}}
+ \bar{P}_{1,L} \left(C_L^{\overline{s}d,s\overline{d}} + C_R^{\overline{s}d,s\overline{d}}\right) \right] ,
	\end{eqnarray}

\noindent where $F_K = 160 $ MeV, $M_K = 497{.}6$ MeV, $ \eta_2 = 0{.}57, \;\; {B}_K = 0{.}85\pm 0{.}15,\;\; \bar{P}_{2,LR} = 30{.}6$
and $\bar{P}_{1,L} = -9{.}3{.}$, we have

	\begin{eqnarray}
\nonumber M_{12}^K &=& \frac{4}{3} F^2_K \eta_2 \bar{B}_K \big(m_d m_s\big) \frac{1}{(v cos\beta )^2} \\
&\times& \sum_{a = 1}^2 \bigg[
 \bar{P}_{2,LR} \frac{U_{2a} U_{1a}}{m_{H}^2}
                 + \bar{P}_{1,L}\bigg(\frac{U_{2a}^2}{m_{H}^2} + \frac{U_{1a}^2}{m_{H}^2}\bigg)
\bigg].
	\end{eqnarray} \\

\subsubsection{ $l_i \to l_j l_k \bar l_k$}\label{sec:Lito3Lj}
\begin{figure}[H]
\centering
\includegraphics[scale = 0.33]{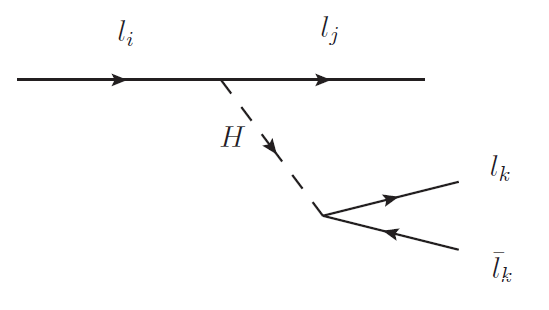}
\caption{ Feynman diagram for the $l_i \to l_j l_k \bar l_k$ process.
\label{fd:liljlklk}}
\end{figure}

The Feynman diagram for the process $l_i \to l_j l_k \bar l_k$ is shown in the FIG.[\ref{fd:liljlklk}],

The Branching ratio for this process including contributions from the three Higgs bosons is given by \cite{lito3lj}

\begin{eqnarray*}
Br(l_{i} & \rightarrow & l_{j}l_{k}\overline{l}_{k})=\frac{5\delta_{l_{l}l_{k}}+2}{3}\frac{\tau_{i}}{2^{11}\pi^{3}}\frac{m_{l_{j}}m_{l_{k}}^{2}m_{l_{i}}^{6}}{v^{6}}\left\{ \frac{cos^{2}(\alpha-\beta)sin^{2}\alpha}{m_{h^{0}}^{4}}\right.\\ \nn
 & + & \frac{sin^{2}(\alpha-\beta)cos^{2}\alpha}{m_{H^{0}}^{4}}-2\frac{cos(\alpha-\beta)sin(\alpha-\beta)\,cos\alpha\, sin\alpha}{m_{h^{0}}^{2}m_{H^{0}}^{2}}\\ \nn
 & + & \left.\frac{sin^{2}\beta}{m_{A^{0}}^{4}}\right\} \frac{\left|\eta_{{ij}}\right|^{2}}{2cos^{4}\beta}.
\end{eqnarray*}

 where $\tau_{i}$ is the time life of the $l_i$ particle. The current upper bounds from PDG \cite{PDG} are:

 \begin{eqnarray*}
BR(\tau\rightarrow\mu\bar{\mu}\mu) & < & 1{.}0\times10^{-12}\\
BR(\mu\rightarrow e\bar{e}e) & < & 2{.}1\times10^{-8}
\end{eqnarray*}

\subsection{Low energy constraints mediated by Charged Higgs bosons}\label{cargados}
\subsubsection{$b \to s \gamma$}\label{bsgamma}

In the Standard Model the dominant decay of b quarks are given via couplings to the $W^{-}$. And rare decays like $b\to s \gamma$ are described by penguin diagrams involving $W^{-}$ at one loop level, but in the THDM heavy charged Higgses also contribute at the one loop-level so the decay $b\to s \gamma$ remains as a good candidate to probe BSM physics.

\begin{figure}[H]
\centering
\includegraphics[width = 0.3\textwidth]{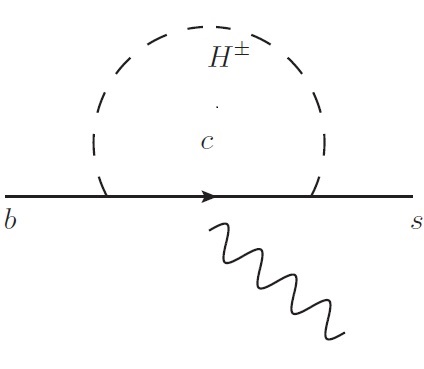}
\caption{ Feynman diagram for the $b \to s \gamma$.
\label{fd:b_sgama}}
\end{figure}

The Feynman diagram for $b\to s \gamma$ involving $H^-$ is shown in FIG.[\ref{fd:b_sgama}].

The decay width is given by

\begin{eqnarray}
\Gamma (b\to s\gamma) = \frac{G_F^2}{32\pi^4}|V_{ts}^\star V_{tb}|^2\alpha_{em}m_b^5\big|\bar D\big|^2
\end{eqnarray}

\noindent where the reduced amplitude $\bar D$ and the Wilson coefficients are presented in
\cite{Borzumati:1998tg}, and discussed in \cite{Ciafaloni:1997un}.

 \subsubsection{$B\rightarrow D(D^*)\tau\nu $}\label{Ds}

The $BABAR$ collaboration \cite{Lees:2013uzd} measured the branching ratios of the $B$ decays $BR(B\to D (D^*)\tau\nu)$, finding:
\begin{eqnarray}
R(D) &=& 0{.}44 \pm 0{.}058 \pm 0{.}042 \\ \nonumber
R(D^*) &=& 0{.}332 \pm 0{.}024 \pm 0{.}018. \nonumber
\end{eqnarray}
where
\begin{equation}
R(D (D^*)) =\frac{ BR(B\to D (D^*)\tau\nu)}{ BR(B\to D (D^*) l \nu)},
\end{equation}
with $l=e,\,\mu$, while the SM theoretical prediction for these processes is given by
\begin{eqnarray}
R_{SM}(D) &=& 0{.}297 \pm 0{.}07, \\ \nonumber
R_{SM}(D^*) &=& 0{.}252 \pm 0{.}003 \pm 0{.}018, \nonumber
\end{eqnarray}
taken together the results give a discrepancy of 3{.}4 $\sigma$. These processes are present in other models that involve charged scalar bosons and can be used to constrain the parameters of such models, like in our case of the THDM-III.

\begin{figure}[H]
\centering
\includegraphics[scale = 0.6]{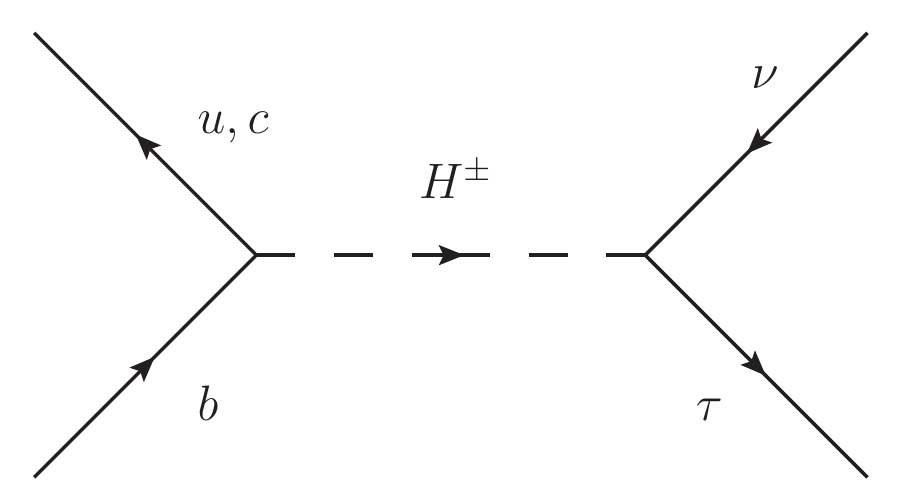}
\caption{ Feynman diagram for the $B\rightarrow D(D^*)\tau\nu$ process.
\label{fd-u_c_b-nu_tau}}
\end{figure}

In this model the Feynman diagram that contributes to this process is shown in FIG.[\ref{fd-u_c_b-nu_tau}]

These ratios can be expressed as follows,

\begin{eqnarray}
{R}(D) &=& {R}_{ SM}(D)\\  &\times & \left(1+ 1{.}5 \;{\mbox{Re}}\left[\frac{C_{R}^{cb}+C_L^{cb}}{C_{SM}^{cb}}\right] + 1{.}0 \left|\frac{C_{R}^{cb}+C_L^{cb}}{C_{SM}^{cb}}\right|^2  \right),\nn
\end{eqnarray}{}

\begin{eqnarray}
{R}(D^*) &=& {R}_{ SM}(D^*)\\ \nn &\times& \left(1 + 0{.}12\; {\mbox{Re}} \left[\frac{C_{R}^{cb}-C_L^{cb}}{C_{SM}^{cb}}\right]+0{.}05 \left|\frac{C_{R}^{cb}-C_L^{cb}}{C_{SM}^{cb}}\right|^2  \right),
\end{eqnarray}

\noindent where $C_{L,R}^{cb}$ are the Wilson coefficients given in \cite{Crivellin:2012ye}.

We find that this is the single most stringent  low energy processes.
Nonetheless, we are able to satisfy both $R(D)$ and $R(D^\star)$ within our model. The combined allowed region of charged Higgs boson mass ($m_H^{\pm}$) and $t_{\beta}$ for all the previous low energy processes are shown in the following section.
\subsection{$h^0\to\tau\mu$}
\begin{figure}[H]
\centering
       \subfigure[ ]{\includegraphics[scale = 0.4]{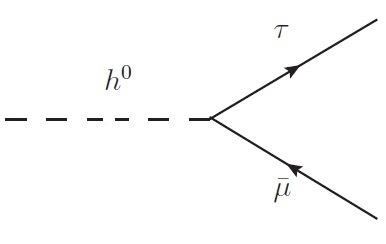}}
\caption{Feynman diagram of $h^0 \to \tau \mu$.\label{htaumu}}
\end{figure}

The flavour violating decay $h^0\to\tau\mu$ currently has an excess of signal events with a significance of 2.4 $\sigma$ \cite{CHMhtaumu}, whose branching ratio reported by CMS with a 95\% confidence level is given by:
\begin{equation}
BR(h^0\to\tau\mu)<1.51\%.
\end{equation}
The tree level Feynman diagram for this process is shown in FIG.[\ref{htaumu}].

\begin{figure}[H]
\centering
       \subfigure[ ]{\includegraphics[scale = 0.3]{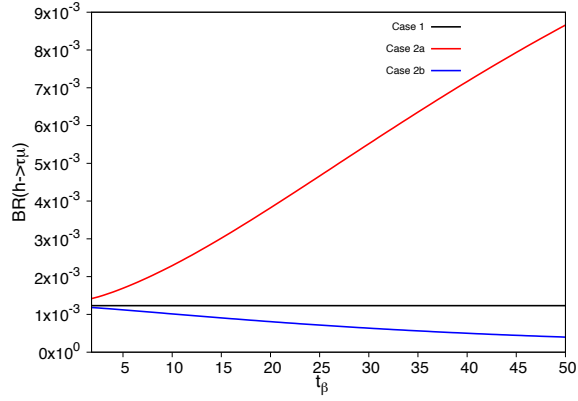}}
\caption{Branching ratio of $h^0\to\tau\mu$ decay as a function of $t_{\beta}$.\label{plothtuamu}}
\end{figure}

It is important to determine if our model acomodates this anomaly, to this purpuse we calculated the decay width given by:
\begin{eqnarray}
\nonumber \Gamma(h^0 \to \tau \mu) &=& \frac{ N_c}{8\pi m_{h^0}^2} \Big|\eta_{\tau\mu}^{h^0}\Big|^2  \left(m_{h^0}^2 -(m_\tau + m_\mu)^2  \right)^{3/2} \\
 &\times& \left(m_{h^0}^2 -(m_\tau - m_\mu)^2  \right)^{1/2},
\end{eqnarray}
\noindent  where $N_c=1$ is the color number and $m_{h^0},\,m_{\tau},\,m_{\mu}$ are the neutral light Higgs boson mass, tau mass and muon mass,  respectively. We found that the Branching ratio for this process as a function of $t_{\beta}$, falls within the LHC reported measurement as shown in FIG.[\ref{plothtuamu}],
  
It might seem that the branching ratio for case 2a could grow in an undesirable way, but it becomes stable at the value of $2\times10^{-2}$, while for the other two cases it decreases as $t_{\beta}$ increases. The allowed range over $t_{\beta}$ that we find in this process matches the allowed range obtained from the low energy process that will be presented in the following section. 

\subsection{Allowed regions}\label{sec:Allowed}

\begin{itemize}
\item {{$t_{\beta}$ and $m_H$ }}{\Large \par}
\end{itemize}
We determine the allowed parameter space by first finding the allowed region for all of the low energy process: \ref{sec:Bto2mu}, \ref{sec:mutoegamma}, \ref{sec:KKmixing},  \ref{sec:Lito3Lj}  and we then intersect this region with the allowed space of $a_{\mu}^{THDM-III}$. In the analysis we have worked with the $\gamma_l=0.5, (\alpha-\beta)=\frac{\pi}{2} $ scenario, $m_{H^{\pm}}=500\,GeV$ $, m_H=m_{H^0}, m_{A^0}$, $\theta_{1,\,2}=0$ and $m_{h^0}=125 \,GeV$; the selection of these values are reasonable, since they satisfy current experimental bounds \cite{haber} on both neutral and charged heavy Higgs.

\begin{figure}[H]
\centering
    \subfigure[]{\includegraphics[width = 0.32\textwidth]{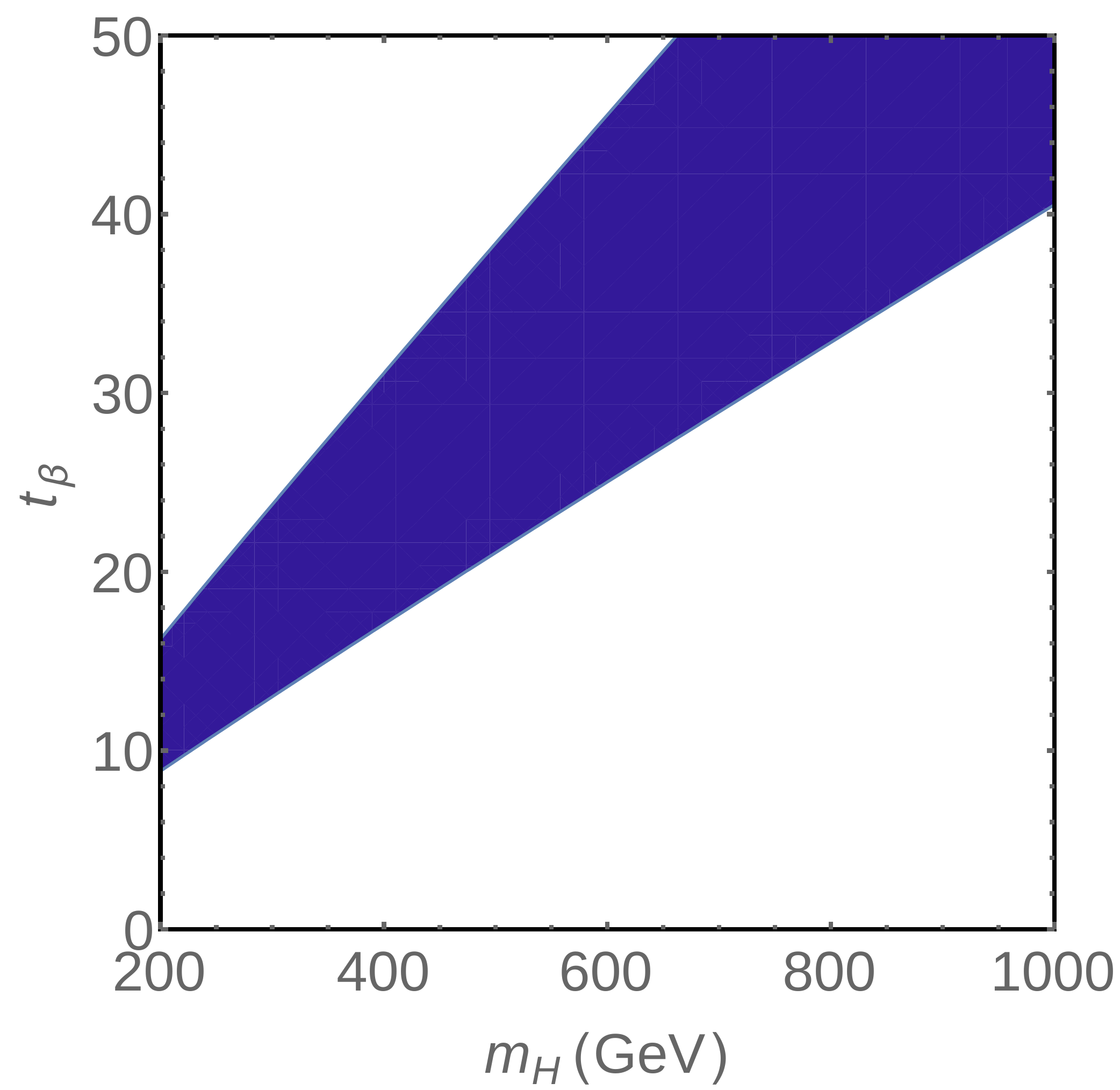}}
     \subfigure[]{\includegraphics[width = 0.32\textwidth]{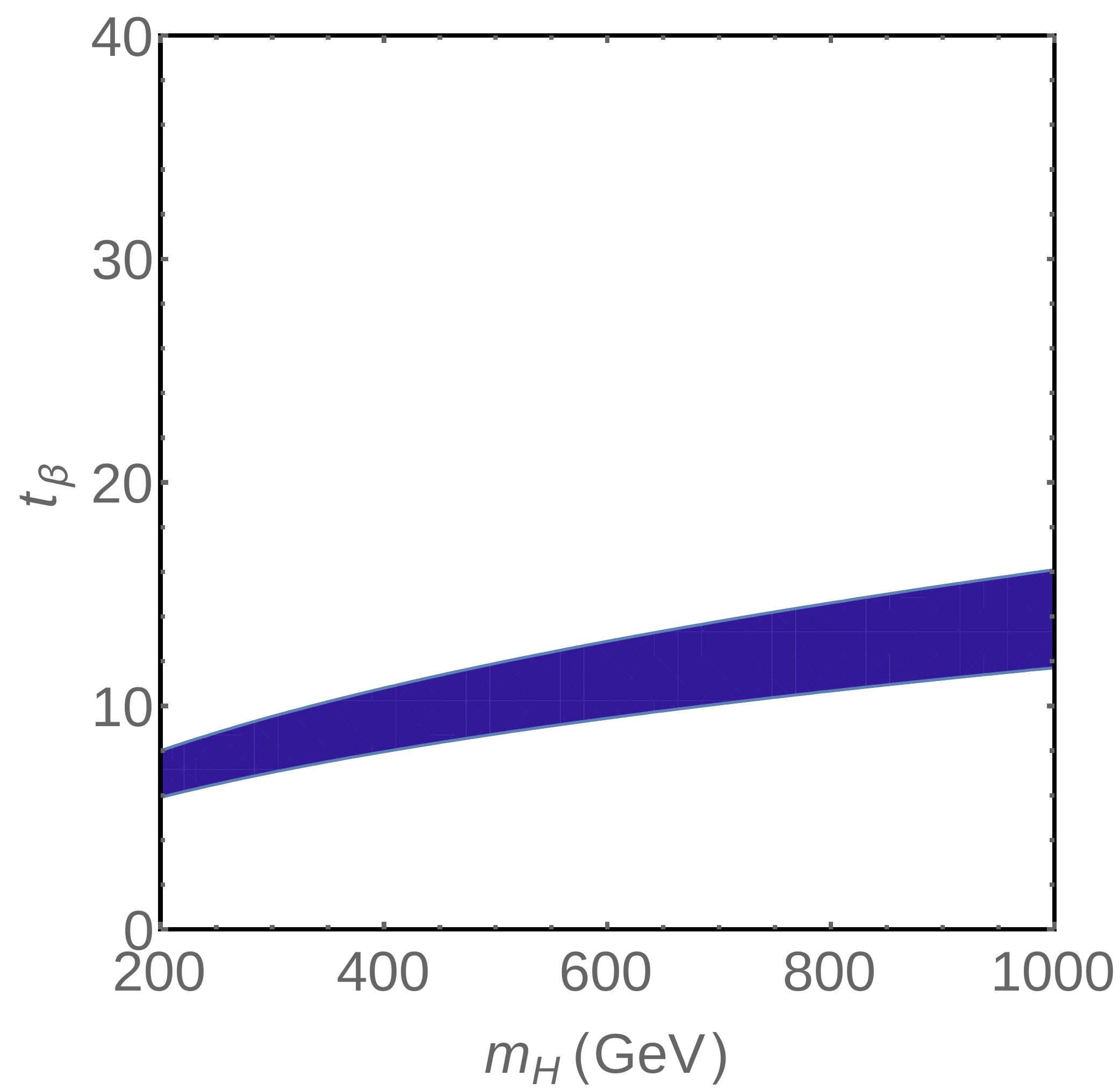}}
      \subfigure[]{\includegraphics[width = 0.32\textwidth]{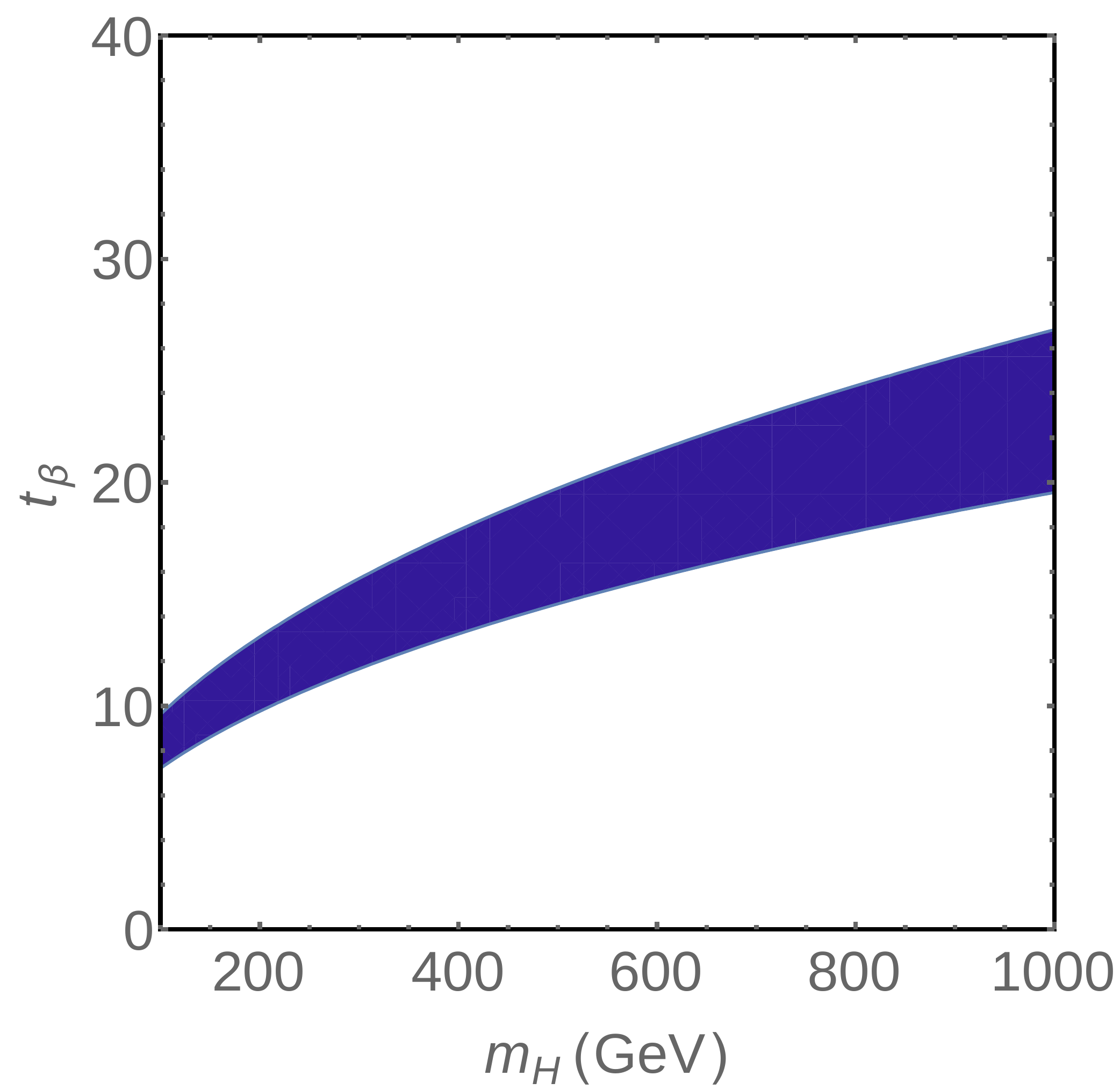}}
     \caption{Allowed regions for the cases: (a) case 1, (b) case 2a, (c) case 2b. \label{neutros}}
	\end{figure}

The allowed regions $(t_{\beta}\,vs\,m_H)$ for case 1, 2a, 2b is presented in FIG.[\ref{neutros}].

Case a) has the largest allowed region, since it allows for small Higgs masses with a small $t_\beta \approx 10$ through heavy Higgs of a large  $t_\beta \approx 50$. While for case 2a we have a much more restricted parameter space, since small masses are allowed for a small range of $t_\beta \approx 6-8$ and the heaviest higgses are between $t_\beta \approx 12-16$. For the case 2b we also see a small allowed zone of parameter space, except that unlike the previous figure, slightly larger values of $t_\beta$ are allowed.

	\begin{itemize}
\item {{$t_{\beta}$ and $m_{H^{\pm}}$ }}{\Large \par}
\end{itemize}
For the charged heavy Higgs bosons we use the constraints \ref{bsgamma}-\ref{Ds} to determine the allowed parameter space  $t_{\beta}\, vs\, m_{H^{\pm}}$. Where we have used  $\gamma=0{.}5, (\alpha-\beta)=\frac{\pi}{2}, m_H=500\,GeV$, $m_{h^0}=125 \,GeV$ and $\theta_1=\frac{\pi}{2},\,\theta_2=\frac{\pi}{4}$.  The allowed regions $(t_{\beta}\,vs\,m_{H^{\pm}})$ for case 1, 2a, 2b is presented in FIG.[\ref{cargados}].

\begin{figure}[h!]
\centering
    \subfigure[]{\includegraphics[width = 0.32\textwidth]{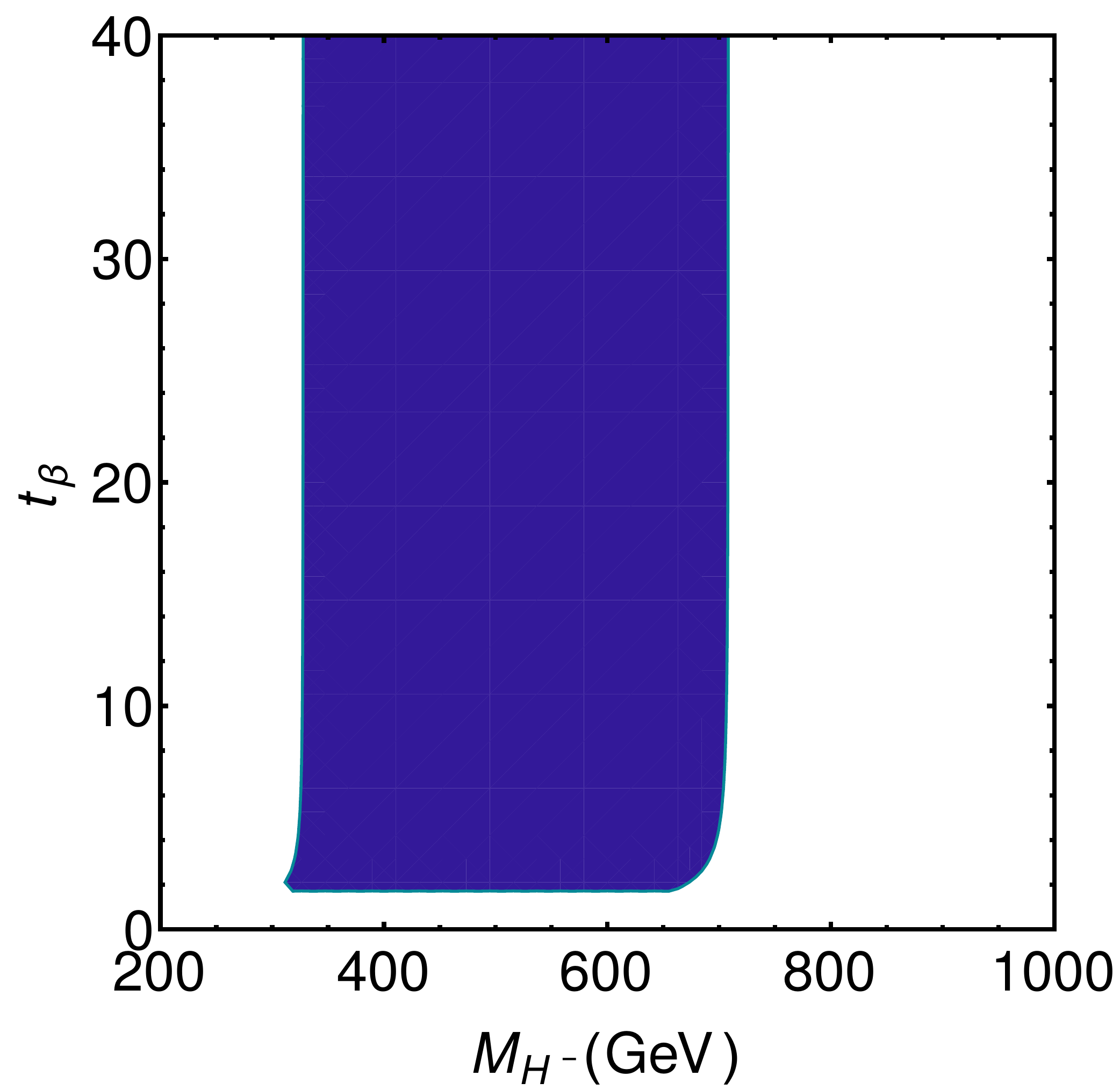}}
     \subfigure[]{\includegraphics[width = 0.32\textwidth]{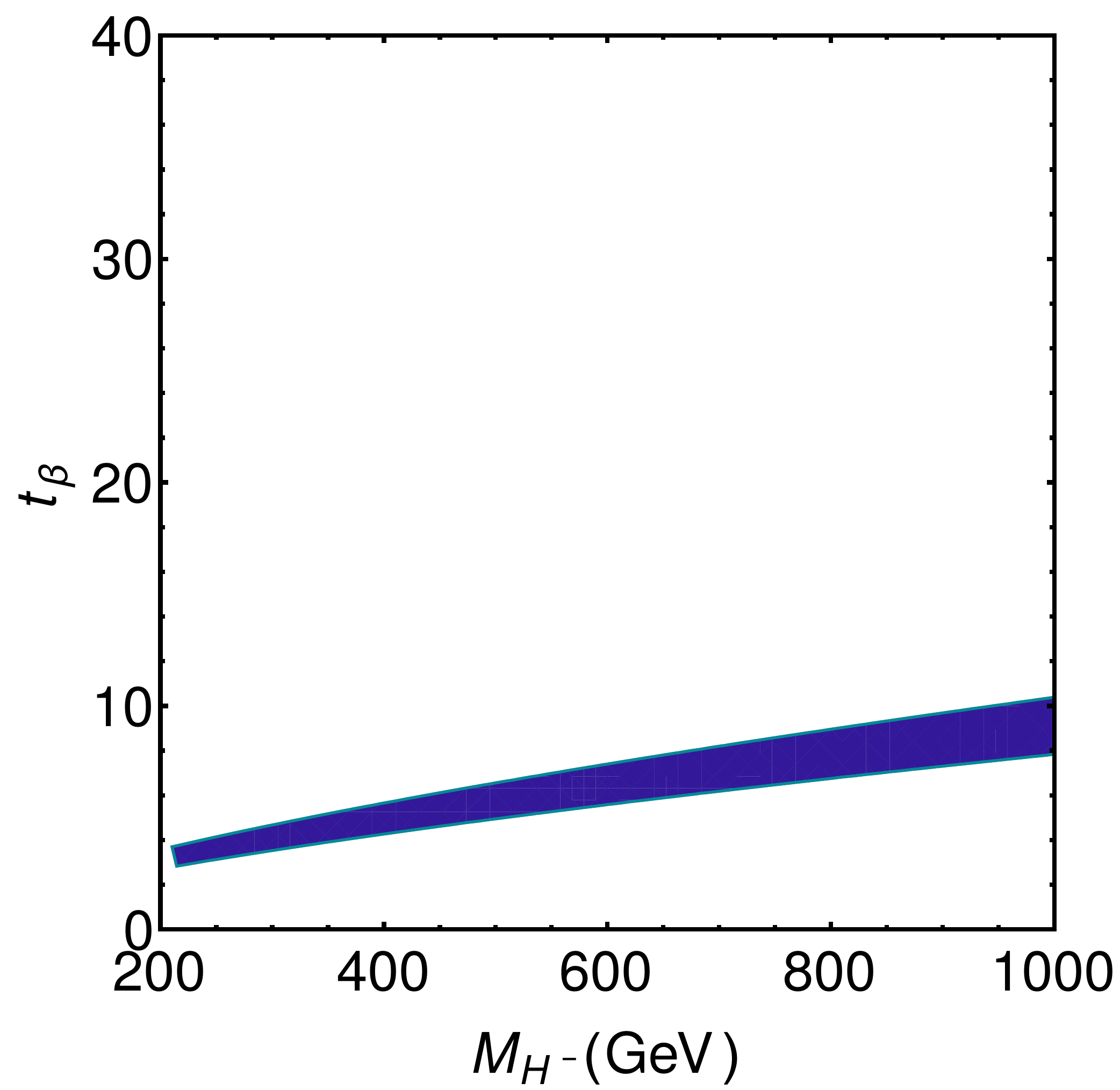}}
      \subfigure[]{\includegraphics[width = 0.32\textwidth]{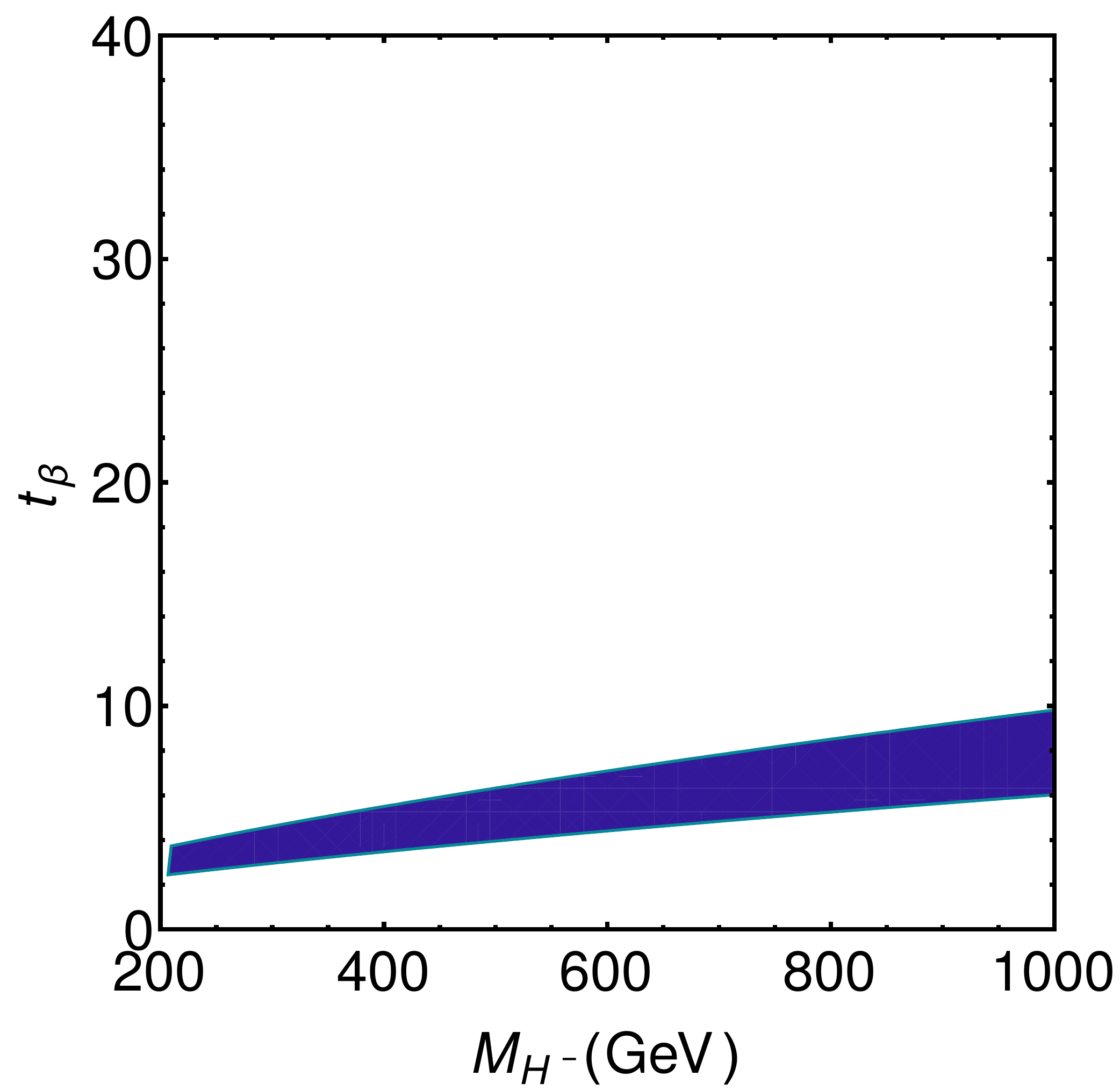}}
     \caption{Allowed regions for the cases: (a) case 1, (b) case 2a, (c) case 2b. \label{cargados}}
	\end{figure}
	
	Constraints on the THDM must also satisfy electroweak precision data (EWPD). Thus, for our results to be reliable we verified that the parameter space regions explored are consistent with EWPD. In particular we found Higgs boson masses and $t_\beta$ that satisfy radiative corrections to the W boson mass in the range of $100<m_H<900\,GeV$ ($m_H=m_{H^0}=m_{A^0}$) and $1<t_{\beta}<40$ for different values of $m_{H^{\pm}}=300,\,500,\,800\,GeV$ and $m_{h^0}=125\,GeV$. Our results agree with the authors of \cite{EWC}.  A brief discution is included in App.\ref{EWdata}.


\section{Results}\label{sec:Numerical}
In this section we show the results obtained for the magnetic and weak magnetic dipole moments in the framework of the THDM-III with four zero textures.
In the previous section we presented the low energy processes involving either charged or neutral Higgs bosons and we explored  what are the available allowed regions that satisfy the experimental bounds or measurements. 

\subsection{Muon anomalous magnetic dipole moment.}

\begin{figure}[H]
\centering
    \subfigure[]{\includegraphics[width = 0.32\textwidth]{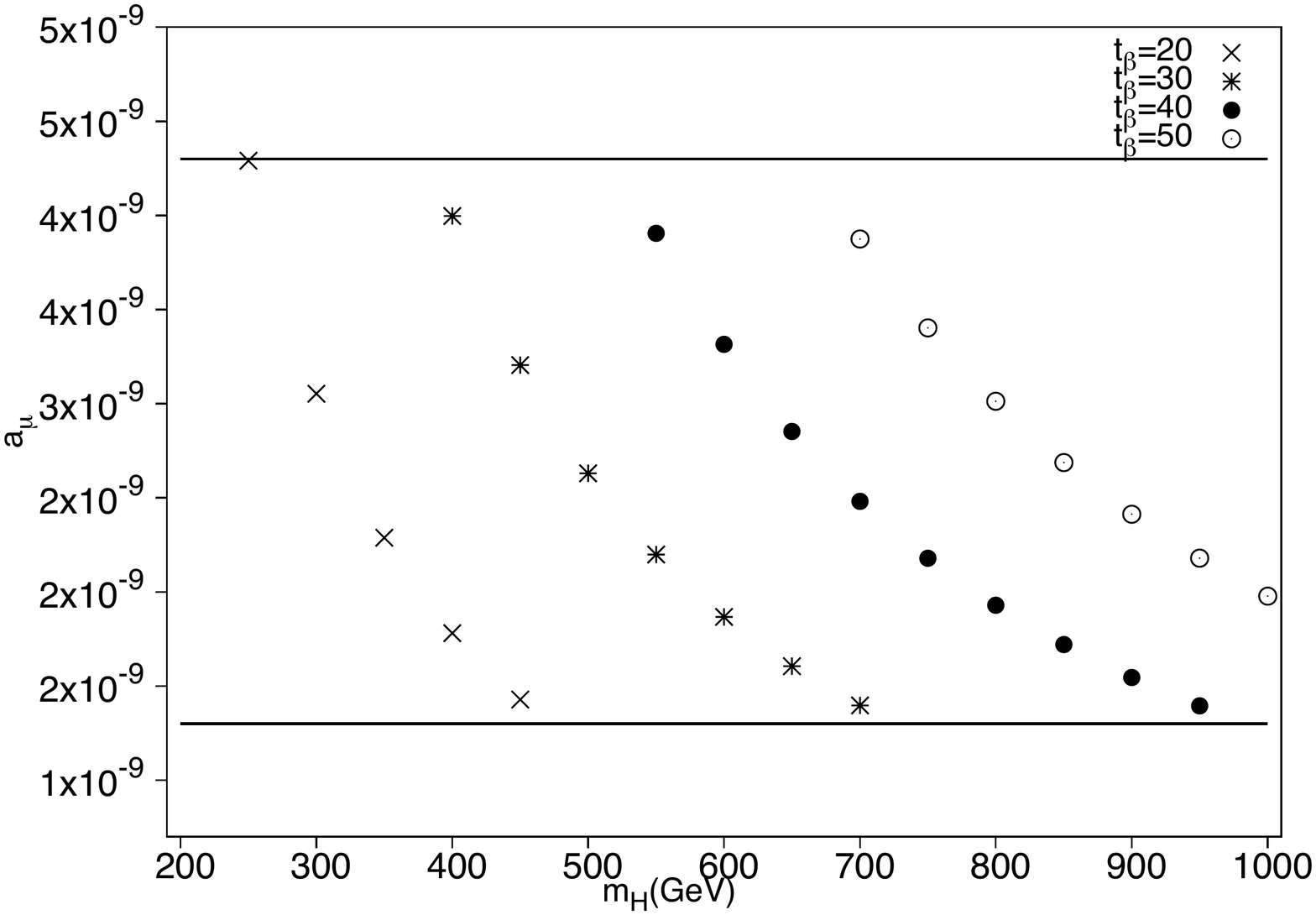}}\label{fig:PMuon}
     \subfigure[]{\includegraphics[width = 0.32\textwidth]{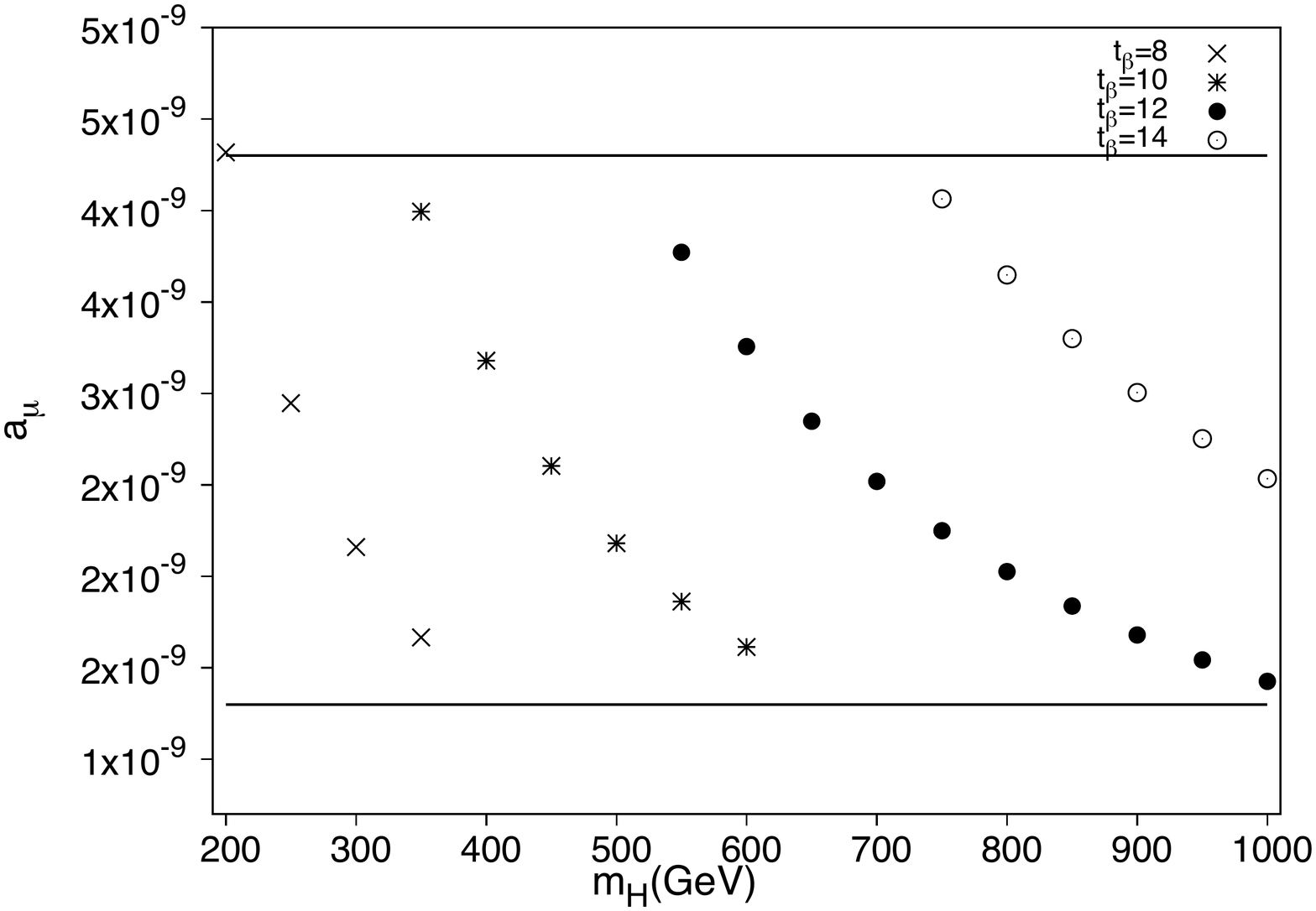}}\label{fig:2acomp}
      \subfigure[]{\includegraphics[width = 0.32\textwidth]{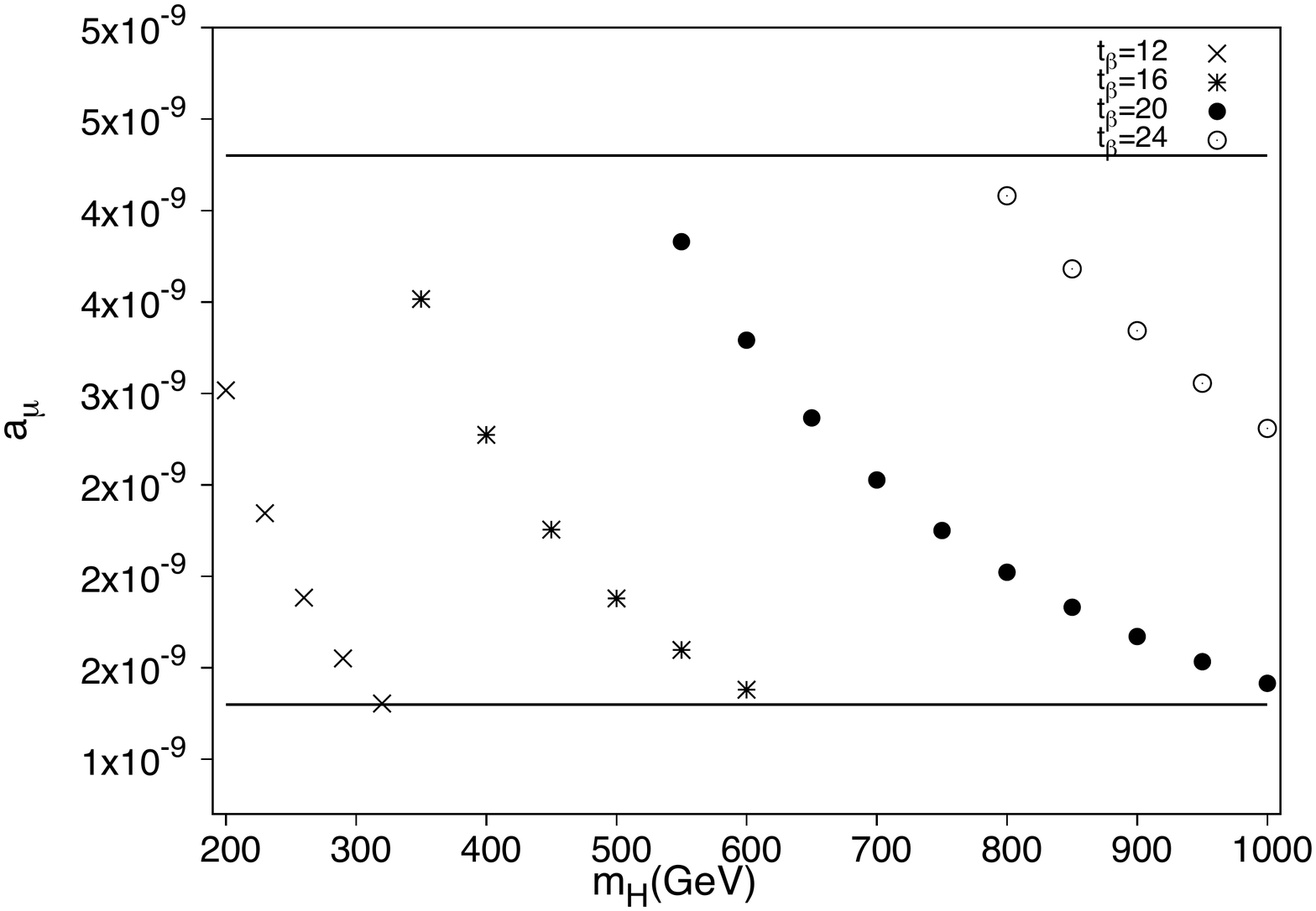}}\label{fig:2bcomp}
     \caption{New contribution of the THDM-III to $a_\mu$ as function of Higgs mass $(m_H)$ and different values of $t_\beta$ for reasonable values of free parameters of the model:   $\gamma=0.5, (\alpha-\beta)=\frac{\pi}{2}, m_H=m_{H^0}= m_{A^0}, m_{H^{\pm}}=500\,GeV$ and $m_{h^0}=125 \,GeV$ for: (a) case 1, (b) case 2a, (c) case 2b. The parallel lines in all the plots correspond to allowed interval of the discrepancy $\Delta a_{\mu}$ between the prediction of the SM and experimental measurements with a 95\% confidence level.\label{fig:Amu}}
	\end{figure}

In FIG.[\ref{fig:Amu}] we show the contributions to $a_\mu$ coming from the one loop diagrams in the context of THDM-III.

In each of the plots of FIG.[\ref{fig:Amu}] we notice that small masses $m_H$ are allowed for small values of $t_\beta$, while heavier Higgs masses are only allowed for greater values of $t_\beta$.

\subsection{Magnetic and weak magnetic dipole moments of the $\tau$.}

\begin{figure}[H]
\centering
    \subfigure[]{\includegraphics[width = 0.32\textwidth]{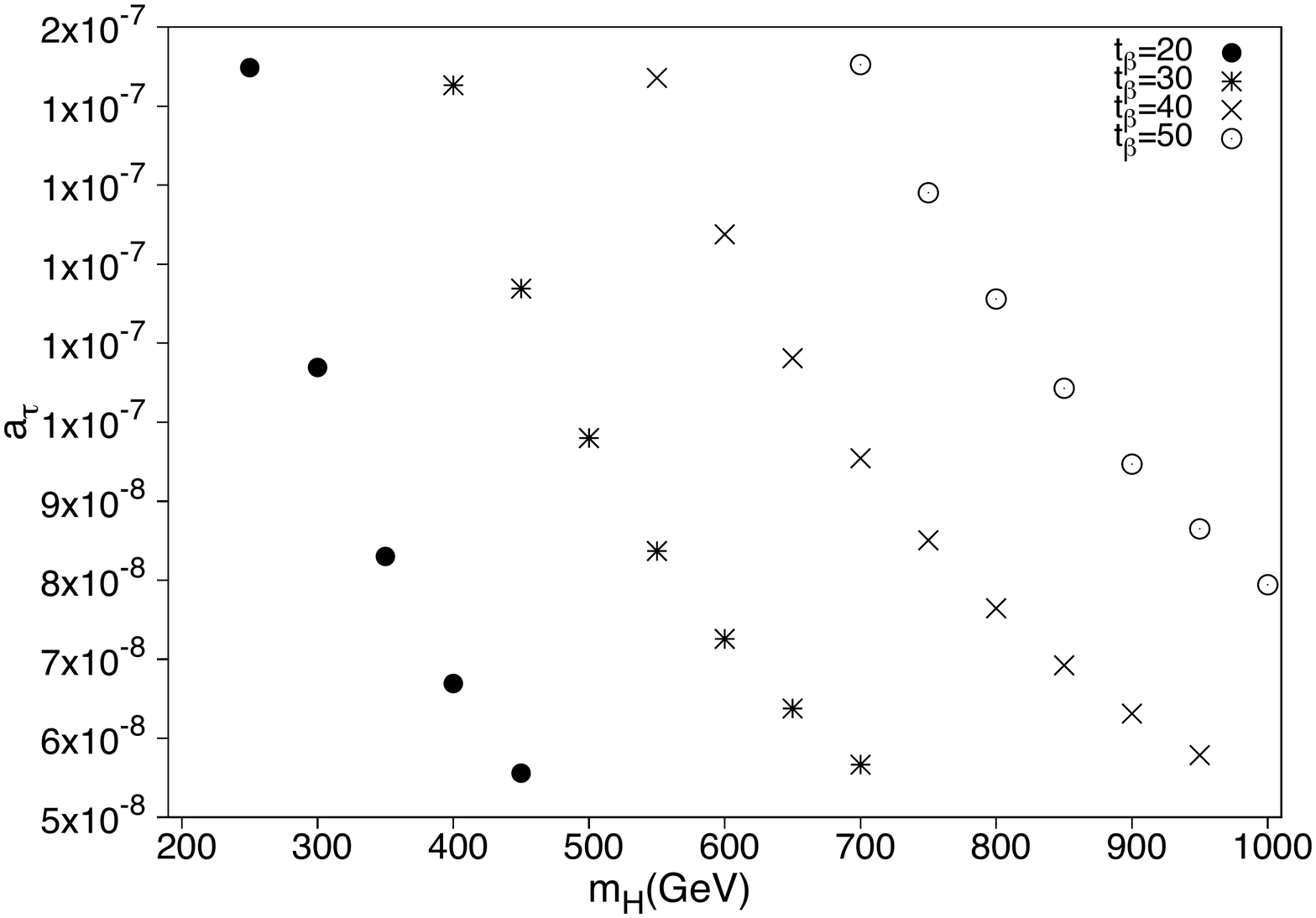}}\label{fig:paraleltau}
    \subfigure[]{\includegraphics[width = 0.32\textwidth]{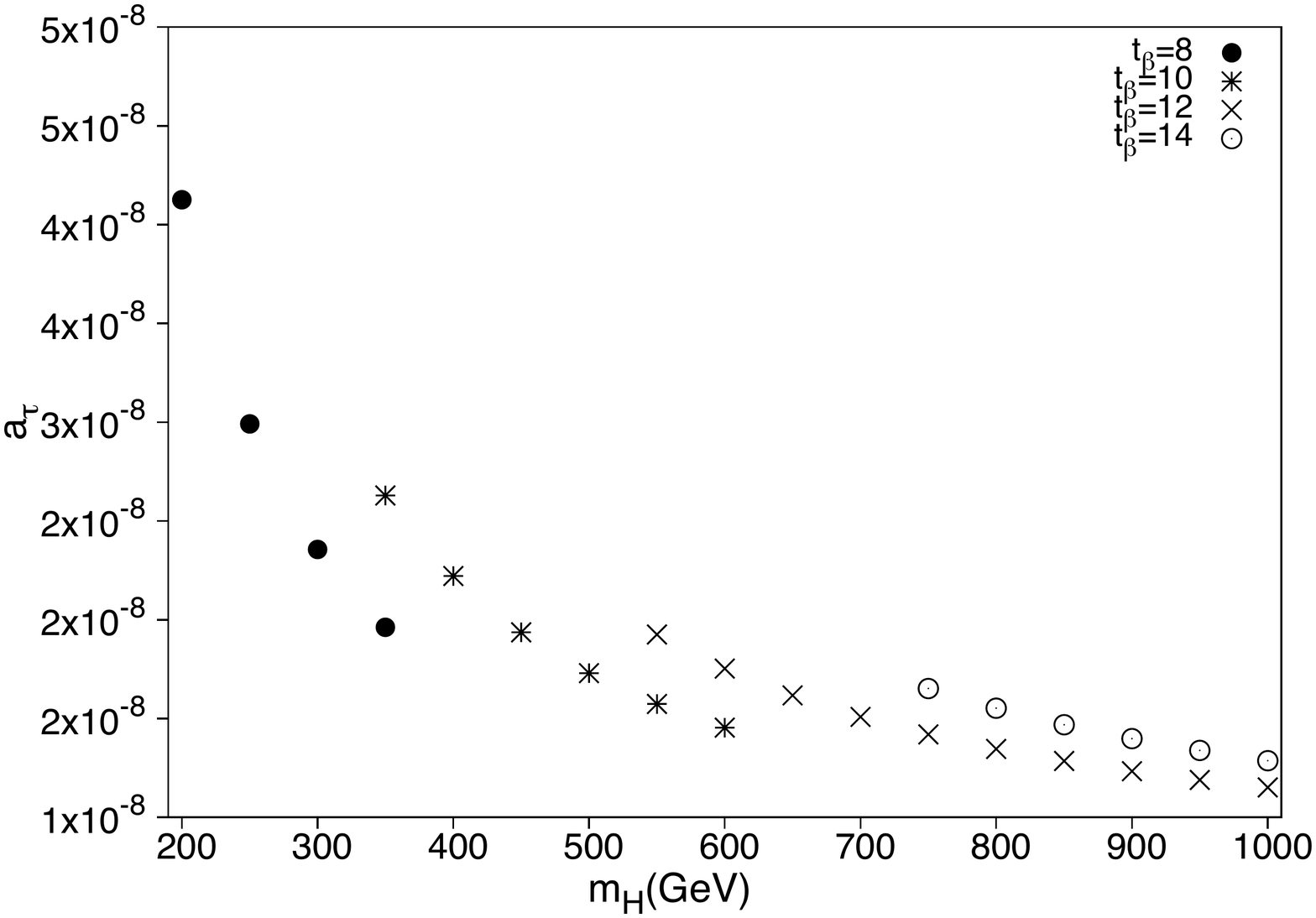}}\label{fig:2atau}
    \subfigure[]{\includegraphics[width = 0.32\textwidth]{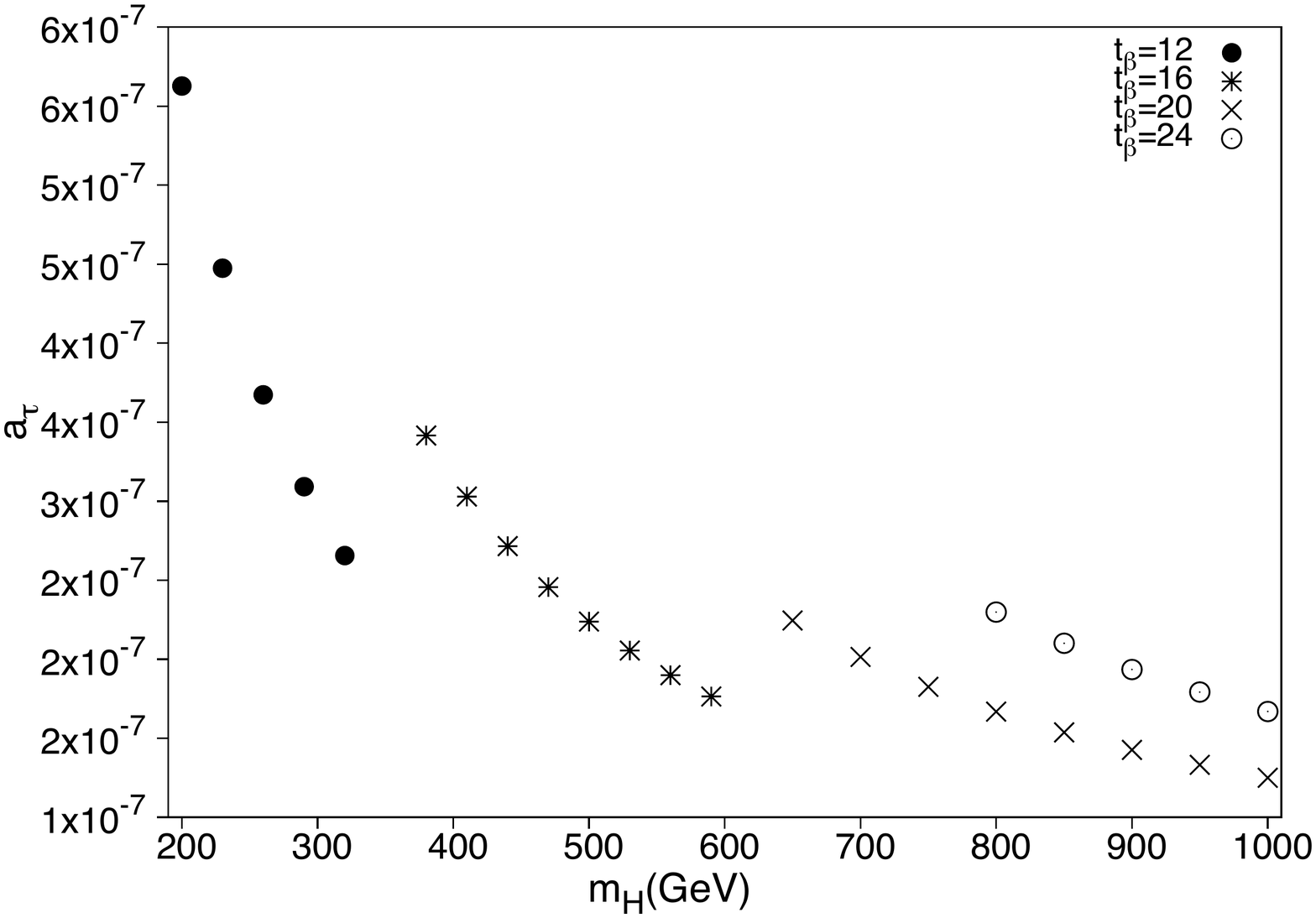}}\label{fig:2btau}
\caption{Plots of the scalar contributions to $a_\tau$ in the THDM-III, (a) case 1, (b) case 2a, (c) case 2b.\label{fig:Atau}}
	\end{figure}

\begin{figure}[H]
\centering
    \subfigure[]{\includegraphics[width = 0.32\textwidth]{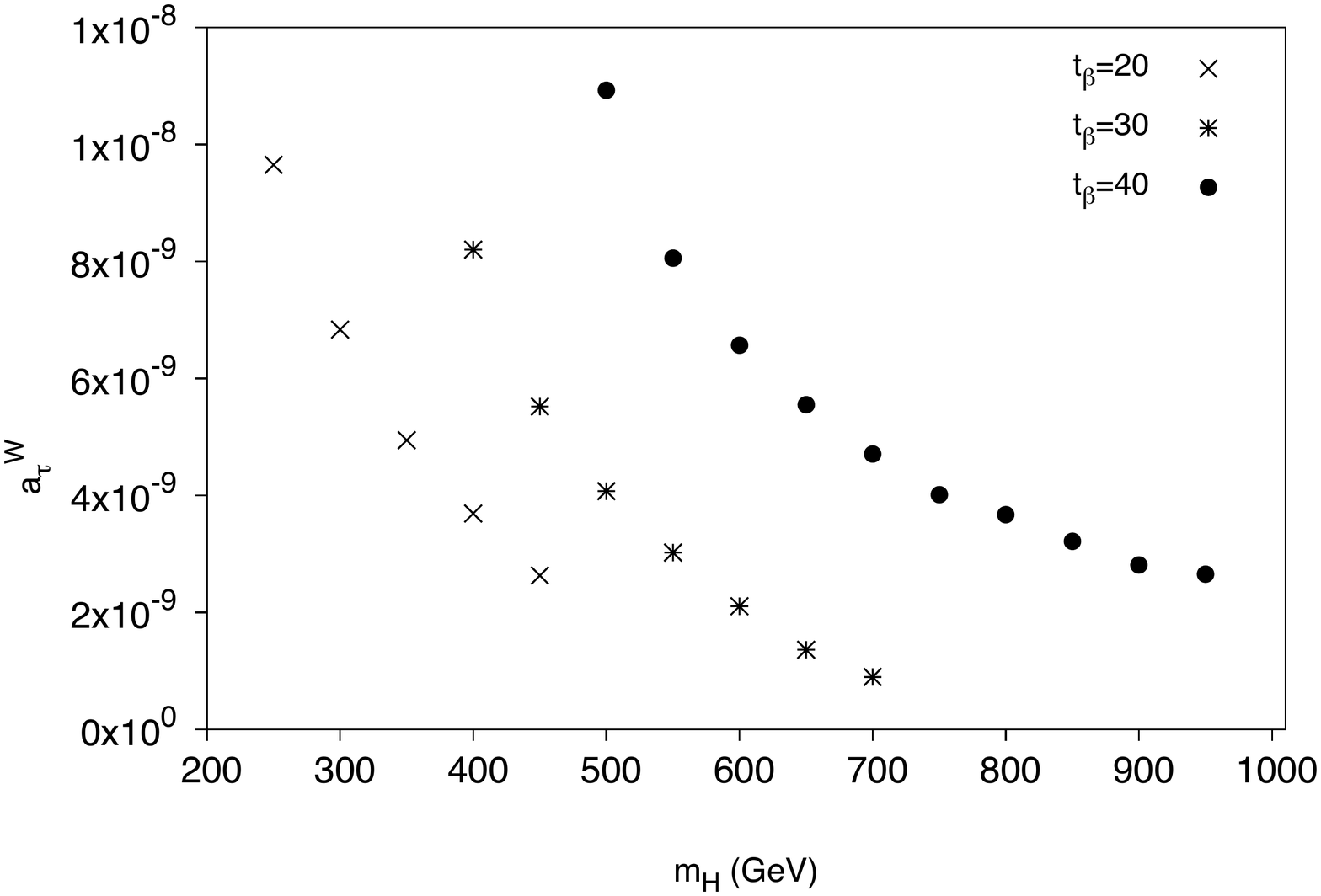}}\label{MomMagParalela.eps}
    \subfigure[]{\includegraphics[width = 0.32\textwidth]{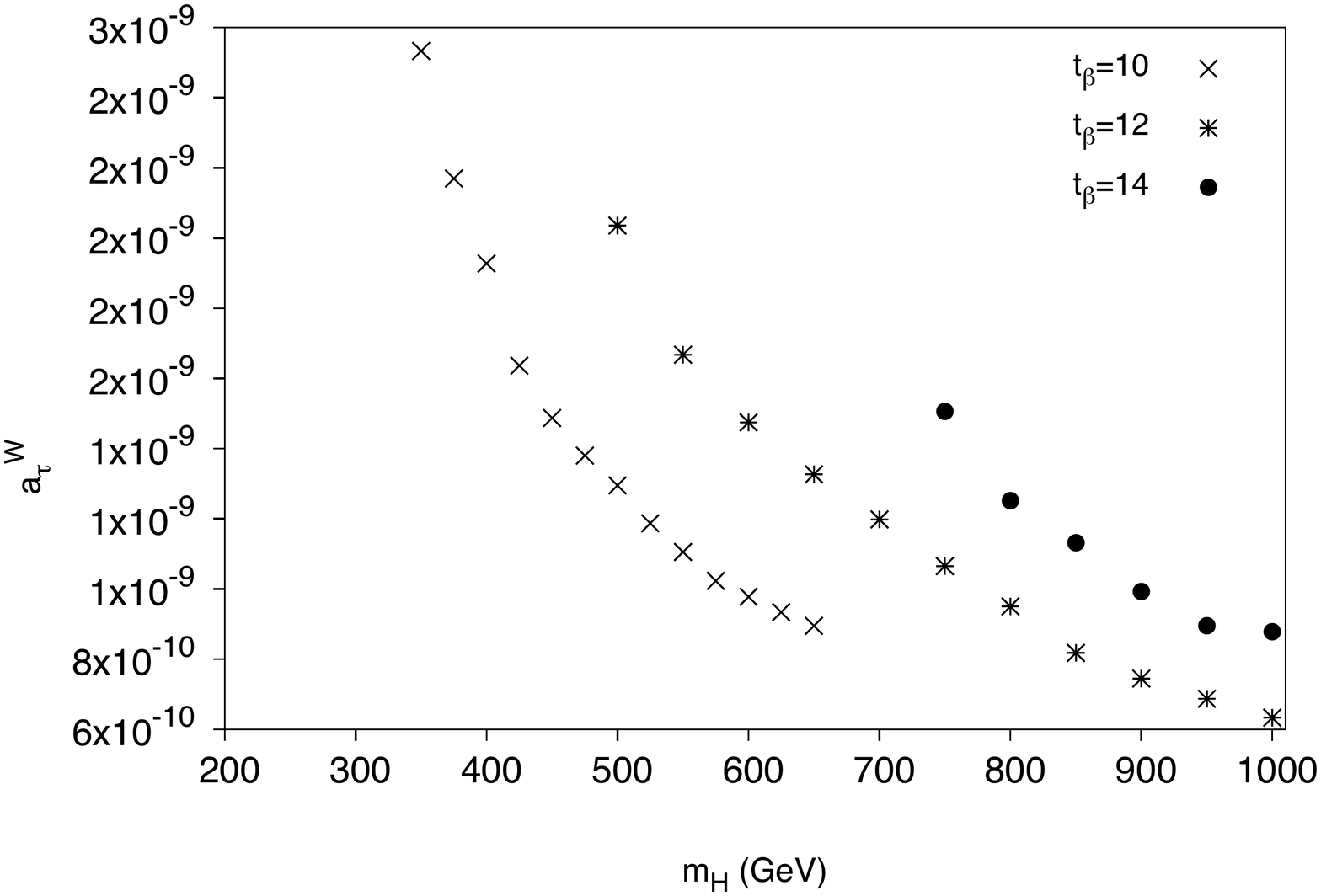}}\label{tauD2a.eps}
    \subfigure[]{\includegraphics[width = 0.32\textwidth]{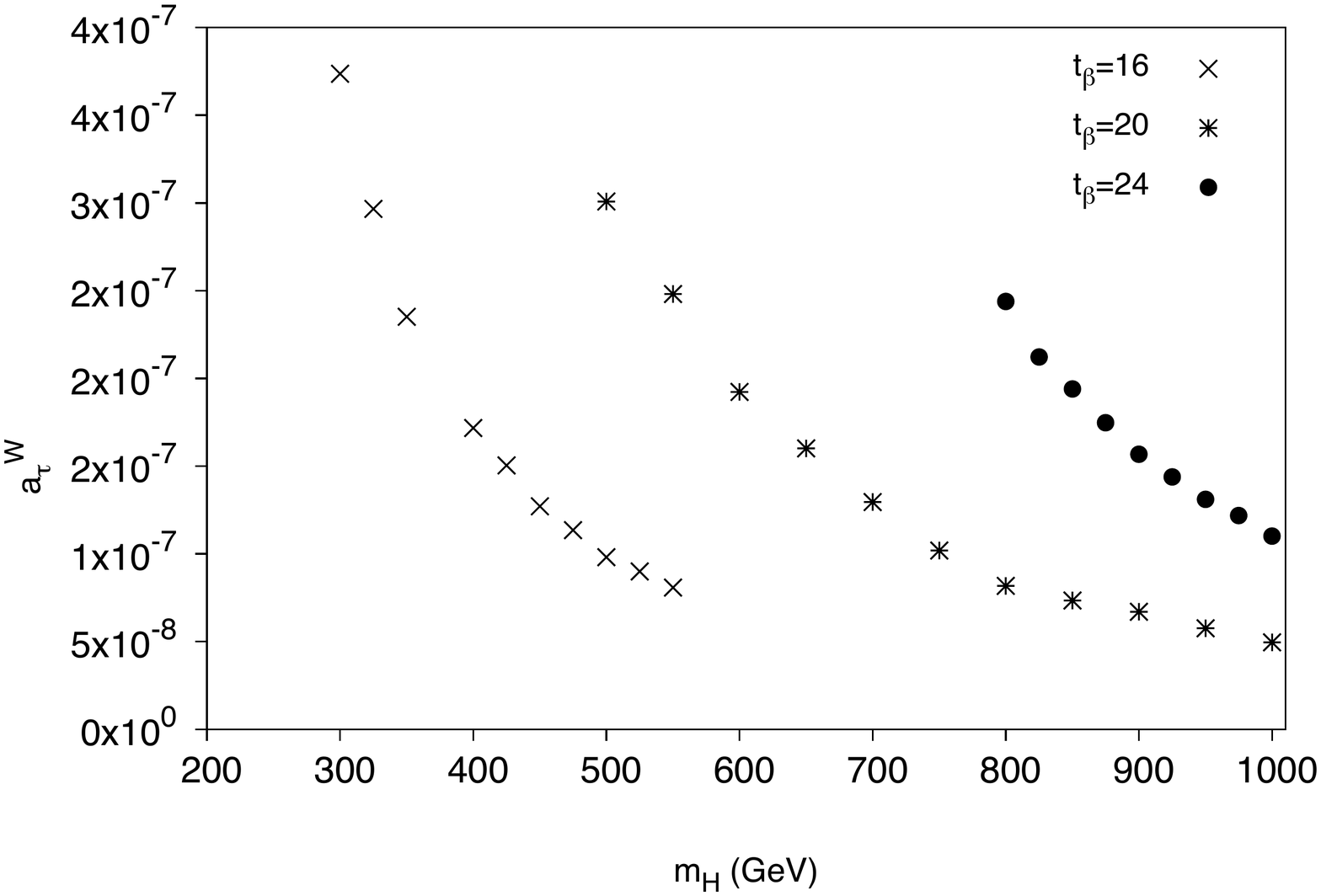}}\label{MomMag2b.eps}
\caption{Plots of the scalar contributions to $a_\tau^W$ in the THDM-III, (a) case 1, (b) case 2a, (c) case 2b.\label{fig:taudebil}}
	\end{figure}

Using the previously obtained parameter space, we show in FIG.[\ref{fig:Atau}] and FIG.[\ref{fig:taudebil}], the THDM-III scalar contributions to the magnetic and weak magnetic dipole moments of the $\tau$.
\noindent from FIG.[\ref{fig:Atau}], we find that the contribution to $a_\tau$  is contained between the current experimental value of \cite{delphi},
\begin{eqnarray}
-0.052 < a_\tau < 0.058.
\end{eqnarray}
We can see that the order of magnitude varies moderately for each of our cases. The contributions of case 1 are within the range $\mathcal{O}(10^{-8}-10^{-7})$ while case 2a and 2b are $\mathcal{O}(10^{-8}),\, \mathcal{O}(10^{-7})$, respectively. It is understandable because the three cases give similar contributions due to the structure of the Yukawa matrices we have considered.

On the other hand, in FIG.[\ref{fig:taudebil}]  shown planes $a_{\tau}^W$ vs $m_H$

We note that the contributions of case 1 are $\mathcal{O}(10^{-9}-10^{-8})$ while case 2a and 2b are $\mathcal{O}(10^{-10}-10^{-9})-\mathcal{O}(10^{-7})$, respectively.

\section{Conclusions}\label{sec:conclusion}

We found that one loop level scalar contributions from the THDM-III with four zero textures can be a possible origin of the discrepancy $\Delta a_\mu$.  For small values of the mass of the heavy Higgs $m_H$,  $t_\beta$ is also small, and reciprocally, larger masses of $m_H$ are permissible only for large $t_\beta$.  Furthermore, using current low energy constraints and $a_\mu^{THDM-III}$, we obtained the allowed parameter space of the Higgs boson masses and $t_\beta$ which are also consistent with electroweak constraints. We determined that the magnetic dipole moment of the $\tau$ lepton is of the order of $\mathcal{O}(10^{-8}-10^{-7})$ and the weak magnetic of $a_{\tau}^W$ is of the order $\mathcal{O}(10^{-10}-10^{-7})$. Of the three cases that we consider the dipole moments found in case 2b have the largest value for $a_{l_i}^{(W)}$.

\begin{acknowledgments}
We acknowledge support from CONACYT (M\'exico). And appreciate the advice and discussions of this work with Gilberto Tavares-Velasco, Lorenzo Diaz-Cruz and Agustin Moyotl.
\end{acknowledgments}



\section{Appendix}

\subsection{Yukawa elements for Case 1} \label{app:chis-cases1-6}

\begin{eqnarray*}
\widetilde{Y}_{21} & = & e^{-i\left(\theta_{1}+\theta_{2}\right)}\sqrt{\frac{\gamma m_{1}m_{2}\left(\gamma m_{2}+m_{2}-m_{3}\right)}{\left(m_{1}+m_{2}\right)\left(m_{1}+m_{3}\right)\left(\gamma m_{2}-m_{3}\right)}}\\
 & \times & \left(e^{i\left(\theta_{1}+\theta_{2}\right)}\sqrt{\frac{\gamma m_{2}^{2}\left(-\gamma m_{2}+m_{1}+m_{3}\right)}{\left(m_{1}+m_{2}\right)\left(m_{2}-m_{3}\right)\left(\gamma m_{2}-m_{3}\right)}}\left(\gamma m_{2}-m_{3}\right)\right.\\
 & + & e^{i\theta_{1}\theta_{2}}\sqrt{\frac{m_{2}\left(\gamma m_{2}+m_{2}-m_{3}\right)}{\left(m_{1}+m_{2}\right)\left(m_{2}-m_{3}\right)}}\\
 & \times & \left.\sqrt{\frac{\gamma m_{2}\left(\gamma m_{2}+m_{2}-m_{3}\right)\left(-\gamma m_{2}+m_{1}+m_{3}\right)}{\gamma m_{2}-m_{3}}}\right)\\
 & + & e^{-i\theta_{1}\theta_{2}}\sqrt{\frac{m_{1}\left(-\gamma m_{2}+m_{1}+m_{3}\right)}{\left(m_{1}+m_{2}\right)\left(m_{1}+m_{3}\right)}}\\
 & \times & \left(-e^{i\theta_{1}\theta_{2}}\sqrt{\frac{m_{2}\left(\gamma m_{2}+m_{2}-m_{3}\right)}{\left(m_{1}+m_{2}\right)\left(m_{2}-m_{3}\right)}}\left(\gamma m_{2}-m_{1}+m_{2}\right)\right.\\
 & - & e^{i\left(\theta_{2}\theta_{1}+\theta_{1}+\theta_{2}\right)}\sqrt{-\frac{m_{1}m_{2}m_{3}}{\gamma m_{2}-m_{3}}}\\
 & \times & \sqrt{-\frac{m_{1}m_{3}\left(\gamma m_{2}+m_{2}-m_{3}\right)}{\left(m_{1}+m_{2}\right)\left(m_{2}-m_{3}\right)\left(\gamma m_{2}-m_{3}\right)}}+e^{i\left(\theta_{1}+\theta_{2}\right)}\\
 & \times & \sqrt{\frac{\gamma m_{2}^{2}\left(-\gamma m_{2}+m_{1}+m_{3}\right)}{\left(m_{1}+m_{2}\right)\left(m_{2}-m_{3}\right)\left(\gamma m_{2}-m_{3}\right)}}\\
 & \times & \left.\sqrt{\frac{\gamma m_{2}\left(\gamma m_{2}+m_{2}-m_{3}\right)\left(-\gamma m_{2}+m_{1}+m_{3}\right)}{\gamma m_{2}-m_{3}}}\right)\\
 & + & e^{-i\left(\theta_{1}+\theta_{2}\right)}\sqrt{\frac{m_{2}\left(\gamma m_{2}+m_{2}-m_{3}\right)}{\left(m_{1}+m_{2}\right)\left(m_{2}-m_{3}\right)}}\sqrt{-\frac{m_{1}m_{2}m_{3}}{\gamma m_{2}-m_{3}}}\\
 & \times & \sqrt{\frac{m_{2}m_{3}\left(\gamma m_{2}-m_{1}-m_{3}\right)}{\left(m_{1}+m_{2}\right)\left(m_{1}+m_{3}\right)\left(\gamma m_{2}-m_{3}\right)}}
\end{eqnarray*}

\begin{eqnarray*}
\widetilde{Y}_{22} & = & 2\sqrt{\frac{m_{2}\left(\gamma m_{2}+m_{2}-m_{3}\right)}{\left(m_{1}+m_{2}\right)\left(m_{2}-m_{3}\right)}}\left(\sqrt{-\frac{m_{1}m_{2}m_{3}}{\gamma m_{2}-m_{3}}}\sqrt{-\frac{m_{1}m_{3}\left(\gamma m_{2}+m_{2}-m_{3}\right)}{\left(m_{1}+m_{2}\right)\left(m_{2}-m_{3}\right)\left(\gamma m_{2}-m_{3}\right)}}\right.\\
 & \times & \cos\left(\theta_{1}+\theta_{2}\right)-\sqrt{\frac{\gamma m_{2}^{2}\left(-\gamma m_{2}+m_{1}+m_{3}\right)}{\left(m_{1}+m_{2}\right)\left(m_{2}-m_{3}\right)\left(\gamma m_{2}-m_{3}\right)}}\\
 & \times & \left.\sqrt{\frac{\gamma m_{2}\left(\gamma m_{2}+m_{2}-m_{3}\right)\left(-\gamma m_{2}+m_{1}+m_{3}\right)}{\gamma m_{2}-m_{3}}}\cos\left(-\theta_{2}\theta_{1}+\theta_{1}+\theta_{2}\right)\right)\\
 & + & \frac{m_{2}\left(m_{2}\left((-2\gamma-1)m_{1}+2\gamma(\gamma+1)m_{2}+m_{2}\right)+m_{3}\left(m_{1}-(2\gamma+1)m_{2}\right)\right)}{\left(m_{1}+m_{2}\right)\left(m_{2}-m_{3}\right)}
\end{eqnarray*}

\begin{eqnarray*}
\widetilde{Y}_{23} & = & e^{-i\left(\theta_{1}+\theta_{2}\right)}\sqrt{-\frac{m_{3}\left(\gamma m_{2}+m_{2}-m_{3}\right)\left(-\gamma m_{2}+m_{1}+m_{3}\right)}{\left(m_{2}-m_{3}\right)\left(m_{1}+m_{3}\right)\left(\gamma m_{2}-m_{3}\right)}}\\
 & \times & \left(e^{i\left(\theta_{1}+\theta_{2}\right)}\sqrt{\frac{\gamma m_{2}^{2}\left(-\gamma m_{2}+m_{1}+m_{3}\right)}{\left(m_{1}+m_{2}\right)\left(m_{2}-m_{3}\right)\left(\gamma m_{2}-m_{3}\right)}}\right.\\
 & \times & \left(\gamma m_{2}-m_{3}\right)+e^{i\theta_{1}\theta_{2}}\sqrt{\frac{m_{2}\left(\gamma m_{2}+m_{2}-m_{3}\right)}{\left(m_{1}+m_{2}\right)\left(m_{2}-m_{3}\right)}}\\
 & \times & \left.\sqrt{\frac{\gamma m_{2}\left(\gamma m_{2}+m_{2}-m_{3}\right)\left(-\gamma m_{2}+m_{1}+m_{3}\right)}{\gamma m_{2}-m_{3}}}\right)\\
 & + & e^{-i\theta_{1}\theta_{2}}\sqrt{-\frac{\gamma m_{2}m_{3}}{\left(m_{2}-m_{3}\right)\left(m_{1}+m_{3}\right)}}\left(e^{i\theta_{1}\theta_{2}}\sqrt{\frac{m_{2}\left(\gamma m_{2}+m_{2}-m_{3}\right)}{\left(m_{1}+m_{2}\right)\left(m_{2}-m_{3}\right)}}\right.\\
 & \times & \left(\gamma m_{2}-m_{1}+m_{2}\right)+e^{i\left(\theta_{2}\theta_{1}+\theta_{1}+\theta_{2}\right)}\sqrt{-\frac{m_{1}m_{2}m_{3}}{\gamma m_{2}-m_{3}}}\\
 & \times & \sqrt{-\frac{m_{1}m_{3}\left(\gamma m_{2}+m_{2}-m_{3}\right)}{\left(m_{1}+m_{2}\right)\left(m_{2}-m_{3}\right)\left(\gamma m_{2}-m_{3}\right)}}\\
 & - & e^{i\left(\theta_{1}+\theta_{2}\right)}\sqrt{\frac{\gamma m_{2}^{2}\left(-\gamma m_{2}+m_{1}+m_{3}\right)}{\left(m_{1}+m_{2}\right)\left(m_{2}-m_{3}\right)\left(\gamma m_{2}-m_{3}\right)}}\\
 & \times & \left.\sqrt{\frac{\gamma m_{2}\left(\gamma m_{2}+m_{2}-m_{3}\right)\left(-\gamma m_{2}+m_{1}+m_{3}\right)}{\gamma m_{2}-m_{3}}}\right)\\
 & + & e^{-i\left(\theta_{1}+\theta_{2}\right)}\sqrt{\frac{m_{2}\left(\gamma m_{2}+m_{2}-m_{3}\right)}{\left(m_{1}+m_{2}\right)\left(m_{2}-m_{3}\right)}}\sqrt{-\frac{m_{1}m_{2}m_{3}}{\gamma m_{2}-m_{3}}}\\
 & \times & \sqrt{\frac{\gamma m_{1}m_{2}^{2}}{\left(m_{2}-m_{3}\right)\left(m_{1}+m_{3}\right)\left(\gamma m_{2}-m_{3}\right)}}
\end{eqnarray*}

\begin{eqnarray*}
\widetilde{Y}_{31} & = & e^{-i\left(\theta_{1}+\theta_{2}\right)}\sqrt{\frac{\gamma m_{1}m_{2}\left(\gamma m_{2}+m_{2}-m_{3}\right)}{\left(m_{1}+m_{2}\right)\left(m_{1}+m_{3}\right)\left(\gamma m_{2}-m_{3}\right)}}\left(e^{i\left(\theta_{1}+\theta_{2}\right)}\right.\\
 & \times & \sqrt{-\frac{m_{3}\left(\gamma m_{2}+m_{2}-m_{3}\right)\left(-\gamma m_{2}+m_{1}+m_{3}\right)}{\left(m_{2}-m_{3}\right)\left(m_{1}+m_{3}\right)\left(\gamma m_{2}-m_{3}\right)}}\left(m_{3}-\gamma m_{2}\right)\\
 & + & e^{i\theta_{1}\theta_{2}}\sqrt{-\frac{\gamma m_{2}m_{3}}{\left(m_{2}-m_{3}\right)\left(m_{1}+m_{3}\right)}}\\
 & \times & \left.\sqrt{\frac{\gamma m_{2}\left(\gamma m_{2}+m_{2}-m_{3}\right)\left(-\gamma m_{2}+m_{1}+m_{3}\right)}{\gamma m_{2}-m_{3}}}\right)\\
 & + & e^{-i\theta_{1}\theta_{2}}\sqrt{\frac{m_{1}\left(-\gamma m_{2}+m_{1}+m_{3}\right)}{\left(m_{1}+m_{2}\right)\left(m_{1}+m_{3}\right)}}\left(-e^{i\theta_{1}\theta_{2}}\sqrt{-\frac{\gamma m_{2}m_{3}}{\left(m_{2}-m_{3}\right)\left(m_{1}+m_{3}\right)}}\right.\\
 & \times & \left(\gamma m_{2}-m_{1}+m_{2}\right)-e^{i\left(\theta_{2}\theta_{1}+\theta_{1}+\theta_{2}\right)}\sqrt{-\frac{m_{1}m_{2}m_{3}}{\gamma m_{2}-m_{3}}}\\
 & \times & \sqrt{\frac{\gamma m_{1}m_{2}^{2}}{\left(m_{2}-m_{3}\right)\left(m_{1}+m_{3}\right)\left(\gamma m_{2}-m_{3}\right)}}\\
 & - & e^{i\left(\theta_{1}+\theta_{2}\right)}\sqrt{\frac{\gamma m_{2}\left(\gamma m_{2}+m_{2}-m_{3}\right)\left(-\gamma m_{2}+m_{1}+m_{3}\right)}{\gamma m_{2}-m_{3}}}\\
 & \times & \left.\sqrt{-\frac{m_{3}\left(\gamma m_{2}+m_{2}-m_{3}\right)\left(-\gamma m_{2}+m_{1}+m_{3}\right)}{\left(m_{2}-m_{3}\right)\left(m_{1}+m_{3}\right)\left(\gamma m_{2}-m_{3}\right)}}\right)\\
 & + & e^{-i\left(\theta_{1}+\theta_{2}\right)}\sqrt{-\frac{m_{1}m_{2}m_{3}}{\gamma m_{2}-m_{3}}}\sqrt{-\frac{\gamma m_{2}m_{3}}{\left(m_{2}-m_{3}\right)\left(m_{1}+m_{3}\right)}}\\
 & \times & \sqrt{\frac{m_{2}m_{3}\left(\gamma m_{2}-m_{1}-m_{3}\right)}{\left(m_{1}+m_{2}\right)\left(m_{1}+m_{3}\right)\left(\gamma m_{2}-m_{3}\right)}}
\end{eqnarray*}

\begin{eqnarray*}
\widetilde{Y}_{32} & = & -e^{i\theta_{1}\theta_{2}}\sqrt{\frac{\gamma m_{2}^{2}\left(-\gamma m_{2}+m_{1}+m_{3}\right)}{\left(m_{1}+m_{2}\right)\left(m_{2}-m_{3}\right)\left(\gamma m_{2}-m_{3}\right)}}\left(e^{-i\theta_{1}\theta_{2}}\right.\\
 & \times & \sqrt{-\frac{m_{3}\left(\gamma m_{2}+m_{2}-m_{3}\right)\left(-\gamma m_{2}+m_{1}+m_{3}\right)}{\left(m_{2}-m_{3}\right)\left(m_{1}+m_{3}\right)\left(\gamma m_{2}-m_{3}\right)}}\left(m_{3}-\gamma m_{2}\right)\\
 & + & e^{-i\left(\theta_{1}+\theta_{2}\right)}\sqrt{-\frac{\gamma m_{2}m_{3}}{\left(m_{2}-m_{3}\right)\left(m_{1}+m_{3}\right)}}\\
 & \times & \left.\sqrt{\frac{\gamma m_{2}\left(\gamma m_{2}+m_{2}-m_{3}\right)\left(-\gamma m_{2}+m_{1}+m_{3}\right)}{\gamma m_{2}-m_{3}}}\right)\\
 & + & e^{-i\theta_{1}\theta_{2}}\sqrt{\frac{m_{2}\left(\gamma m_{2}+m_{2}-m_{3}\right)}{\left(m_{1}+m_{2}\right)\left(m_{2}-m_{3}\right)}}\left(e^{i\theta_{1}\theta_{2}}\sqrt{-\frac{\gamma m_{2}m_{3}}{\left(m_{2}-m_{3}\right)\left(m_{1}+m_{3}\right)}}\right.\\
 & \times & \left(\gamma m_{2}-m_{1}+m_{2}\right)+e^{i\left(\theta_{2}\theta_{1}+\theta_{1}+\theta_{2}\right)}\sqrt{-\frac{m_{1}m_{2}m_{3}}{\gamma m_{2}-m_{3}}}\\
 & \times & \sqrt{\frac{\gamma m_{1}m_{2}^{2}}{\left(m_{2}-m_{3}\right)\left(m_{1}+m_{3}\right)\left(\gamma m_{2}-m_{3}\right)}}\\
 & + & e^{i\left(\theta_{1}+\theta_{2}\right)}\sqrt{\frac{\gamma m_{2}\left(\gamma m_{2}+m_{2}-m_{3}\right)\left(-\gamma m_{2}+m_{1}+m_{3}\right)}{\gamma m_{2}-m_{3}}}\\
 & \times & \left.\sqrt{-\frac{m_{3}\left(\gamma m_{2}+m_{2}-m_{3}\right)\left(-\gamma m_{2}+m_{1}+m_{3}\right)}{\left(m_{2}-m_{3}\right)\left(m_{1}+m_{3}\right)\left(\gamma m_{2}-m_{3}\right)}}\right)\\
 & + & e^{-i\left(\theta_{1}+\theta_{2}\right)}\sqrt{-\frac{m_{1}m_{2}m_{3}}{\gamma m_{2}-m_{3}}}\sqrt{-\frac{m_{1}m_{3}\left(\gamma m_{2}+m_{2}-m_{3}\right)}{\left(m_{1}+m_{2}\right)\left(m_{2}-m_{3}\right)\left(\gamma m_{2}-m_{3}\right)}}\\
 & \times & \sqrt{-\frac{\gamma m_{2}m_{3}}{\left(m_{2}-m_{3}\right)\left(m_{1}+m_{3}\right)}}
\end{eqnarray*}

\begin{eqnarray*}
\widetilde{Y}_{33} & = & 2\sqrt{-\frac{\gamma m_{2}m_{3}}{\left(m_{2}-m_{3}\right)\left(m_{1}+m_{3}\right)}}\left(\sqrt{-\frac{m_{1}m_{2}m_{3}}{\gamma m_{2}-m_{3}}}\right.\\
 & \times & \sqrt{\frac{\gamma m_{1}m_{2}^{2}}{\left(m_{2}-m_{3}\right)\left(m_{1}+m_{3}\right)\left(\gamma m_{2}-m_{3}\right)}}\cos\left(\theta_{1}+\theta_{2}\right)\\
 & + & \sqrt{\frac{\gamma m_{2}\left(\gamma m_{2}+m_{2}-m_{3}\right)\left(-\gamma m_{2}+m_{1}+m_{3}\right)}{\gamma m_{2}-m_{3}}}\\
 & \times & \left.\sqrt{-\frac{m_{3}\left(\gamma m_{2}+m_{2}-m_{3}\right)\left(-\gamma m_{2}+m_{1}+m_{3}\right)}{\left(m_{2}-m_{3}\right)\left(m_{1}+m_{3}\right)\left(\gamma m_{2}-m_{3}\right)}}\cos\left(-\theta_{2}\theta_{1}+\theta_{1}+\theta_{2}\right)\right)\\
 & + & \frac{m_{3}\left(m_{3}\left(2\gamma m_{2}-m_{1}+m_{2}\right)+m_{2}\left(2\gamma m_{1}-2\gamma(\gamma+1)m_{2}+m_{1}\right)-m_{3}^{2}\right)}{\left(m_{2}-m_{3}\right)\left(m_{1}+m_{3}\right)}
\end{eqnarray*}

\subsection{Electroweak constraints} \label{EWdata}
Following the work done in \cite{Sola} the W mass in the THDM is perturbatively calculated through the relation 
\begin{equation}
M_{W}^2=\frac{M_{Z}^2}{2}\left(1+\sqrt{1-\frac{4\pi\alpha_{em}}{\sqrt{2}G_F M_{Z}^2}(1+\Delta r)} \right),
\end{equation} 
where $G_F,\,\alpha_{em},\,M_Z$ are the Fermi constant, fine-structure constant and Z boson mass, respectively. The terms $\Delta r$ are radiative corrections to the W boson mass given by:
\begin{equation}
\Delta r=\Delta r^{SM}+\Delta r^{THDM}
\end{equation}  
where
\begin{equation}
\Delta r^{THDM}=\Delta\alpha^{THDM}-\frac{c_W^2}{s_W^2}\delta\rho^{THDM}+\Delta r_{rem},
\end{equation}  
such that $\Delta\alpha^{THDM}$ is the photon vacuum polarization of the THDM, $\Delta r_{rem}$ incorporates the remaining contributions  and
\begin{eqnarray}
\nonumber \delta \rho^{THDM} &=& \frac{-\alpha}{16 \pi s_W^2 M_W^2}  \Big{(} \cos^2 (\beta - \alpha) [ F(m_{h^0}^2, m_{H^{\pm}}^2) - F(m_{h^0}^2, m_{A^0}^2) ] \\
\nonumber &+& \sin^2 (\beta - \alpha) [ F(m_{H^0}^2 , m_{H^{\pm}}^2) - F(m_{H^0}^2, m_{A^0}^2)]  + F(m_{A^0}^2, m_{H^\pm}^2) \\
\nonumber &-& 3\cos^2(\beta - \alpha) [ F(m_{H^0}^2 , M_{W}^2) + F(m_{h^0}^2 , M_{Z}^2) ] \\
&-& F(m_{H^0}^2 , M_{Z}^2) - F(m_{h^0}^2 , M_{W}^2)\Big{)},
\end{eqnarray}  

\begin{eqnarray}
F(x,\,y)&=&\frac{x+y}{2}-\frac{xy}{x-y} log(\frac{x}{y}),\, if\, x\neq y \\
F(x,\,y)&=&0,\,if\, x = y.
\end{eqnarray}


\begin{thebibliography}{99}

\bibitem{MomMagMuonEXP}
G. Bennett et al. (Muon G-2 Collaboration), Phys.Rev.D73, 072003 (2006), arXiv:hep-ex/0602035.
\bibitem{Pich}
  M. Jung and A. Pich, JHEP1404, 076 (2014) [arXiv:1308.6283 [hep-ph].

\bibitem{Dumm}
 D. Gomez-Dumm and G. Gonzalez-Sprinberg, Eur.Phys.J.C11, 293 (1999), arXiv:hep-ph/9905213 [hep-ph].

\bibitem{delphi}
J. Abdallah et al. [DELPHI Collaboration], Eur. Phys. J. C 35 (2004) 159.

\bibitem{MomMagtauSM}
S. Eidelman and M. Passera, Mod.Phys.Lett. A22, 159 (2007), arXiv:hep-ph/0701260 [hep-ph].

 \bibitem{LQ}
  Bola\~nos, A. and Moyotl, A. and Tavares-Velasco, G., Phys. Rev. D89 2014, arxiv:1312.6860

 \bibitem{MSSM1}
  T. Ibrahim and P. Nath, Phys.Rev. D78 , 075013 (2008), arXiv:0806.3880 [hep-ph].
  
 \bibitem{UP}
 A. Moyotl and G. Tavares-Velasco, Phys.Rev.D86, 013014 (2012), arXiv:1210.1994 [hep-ph]
 
\bibitem{ValorSM-MDMDtau}
J. Bernabeu, G. Gonzalez-Sprinberg, M. Tung, and J. Vidal, Nucl.Phys.B436, 474 (1995), arXiv:hep-ph/9411289 [hep-ph]





 \bibitem{MSSM3}
  B. de Carlos and J. Moreno, Nucl.Phys. B519 , 101 (1998), arXiv:hep-ph/9707487 [hep-ph]; W. Hollik, J. I. Illana,
S. Rigolin, and D. Stockinger, Phys.Lett. B416 , 345 (1998), arXiv:hep-ph/9707437 [hep-ph].


 
\bibitem{MOMWEAK-EXP}
A. Heister et al. (ALEPH Collaboration), Eur.Phys.J. C30, 291 (2003), arXiv:hep-ex/0209066.




\bibitem{Tavares12}
   M. Laursen, M. A. Samuel, and A. Sen, Phys.Rev.D29, 2652 (1984).

\bibitem{chinos}
Qing-Jun Xu, Chao-Hsi Chang, arXiv:1410.1774v2 [hep-ph]

\bibitem{Buras:2010mh}
Buras, Andrzej J. and Carlucci, Maria Valentina and Gori, Stefania and Isidori, Gino.
JHEP, \ 10, \ 2010.

\bibitem{Cheng:1987rs} 
  T.~P.~Cheng and M.~Sher,
  Phys.\ Rev.\ D {\bf 35}, 3484 (1987).
  
\bibitem{Honda:2004qh}
Honda, Mizue and Kaneko, Satoru and Tanimoto, Morimitsu,
Phys. Lett.\ B593. \ 2004.

\bibitem{Branco:2010tx}
Branco, G. C. and Emmanuel-Costa, D. and Simoes, C.
Phys. Lett., \ B690, \ 2010.

 \bibitem{BrancoReview}
   G. C. Branco, P. M. Ferreira, L. Lavoura, M. N. Rebelo, M. Sher, and J. P. Silva, Theory  and  phenomenology  of
two-Higgs-doublet models, Phys. Rept. 516(2012) 1 [arXiv:1106.0034 [hep-ph]].

\bibitem{HNishiura}
H. Nishiura, K. Matsuda and T. Fukuyama, PHYSICAL REVIEW D, VOLUME 60, 013006.
\bibitem{papaqui}
J.L. Diaz-Cruz, R. Noriega-Papaqui and A. Rosado,

\bibitem{Branco:1999nb}
  G.~C.~Branco, D.~Emmanuel-Costa and R.~Gonzalez Felipe,
  Phys.\ Lett.\ B {\bf 477}, 147 (2000)
  [hep-ph/9911418].



\bibitem{GomezBock:2005hc}
  M.~Gomez-Bock and R.~Noriega-Papaqui,
  J.\ Phys.\ G {\bf 32}, 761 (2006)
  [hep-ph/0509353].


\bibitem{Gunion:1989we}
  J.~F.~Gunion, H.~E.~Haber, G.~L.~Kane and S.~Dawson,
  ``The Higgs Hunter's Guide,'' Front.\ Phys.\  {\bf 80}, 1 (2000).
  
  \bibitem{Barr-Zee}
  D. Chang, W.-F. Chang, C.-H. Chou, and W.-Y. Keung, Phys.Rev. D 63, 091301(R)(2001).
  \bibitem{TavCP}
  F. Larios, G- Tavares-Velasco, C.-P Yuan, Phys-Rev. D 64, 055004. 
  
  \bibitem{OPAL}
  OPAL Collaboration, G. Abbiendi et.al., Eur. Phys. J. C 18, 425 (2001).
  
  \bibitem{alpha-beta}
  Dean Carmi, Adam Falkowski, Eric Kuflik, Tomer Volansky, Jure Zupan, 10.1007/JHEP10(2012)196, arXiv:1207.1718v3 [hep-ph]. 

\bibitem{mHcota}
J. Abdallah et al. [DELPHI Collaboration], Eur. Phys. J. C 38 (2004) 1 [hep-ex/0410017].



\bibitem{mHcotainferior}
The OPAL Collaboration, Search for a low mass CP-odd Higgs boson in electron positron collisions with the OPAL detector  at LEP2, Eur. Phys. J. C27 (2003), [hep-ex/0209068].



\bibitem{MarcoKikeTHDMTX}
M. Arroyo, J. L. Diaz-Cruz, E. Diaz and J. A Orduz-Ducuara, arXiv:1306.2343 [hep-ph]

\bibitem{RGETHDM}
G. CVETIC, S. S. HWANG, and C. S. KIM, Int. J. Mod. Phys. A 14, 769 (1999). DOI: 10.1142/S0217751X99000385 

\bibitem{RGETHDM4zeros}
Zhi-zhong Xing, Zhen-hua Zhao. DOI:10.1016/j.nuclphysb.2015.05.027.  arXiv:1501.06346v2 [hep-ph].

\bibitem{haber}
P.M. Ferreira, R. Guedes, J.F. Gunion, H.E. Haber, M.O.P. Sampaio, R. Santos, The
Wrong Sign limit in the 2HDM, arXiv:1410.1926 [hep-ph].



\bibitem{ALEPH}
A. Heister et al. (ALEPH Collaboration), Eur.Phys.J. C30, 291 (2003), arXiv:hep-ex/0209066 [hep-ex].

\bibitem{Khachatryan}
V. Khachatryan et al., [CMS and LHCb Collaborations], arXiv:1411.4413 [hep-ex].

\bibitem{BmumuSM}
  C. Bobeth et al. , Phys. Rev. Lett. 112 (2014) 101801.

 \bibitem{Dedes}
A. Dedes and A. Pilaftsis, Phys. Rev. D 67, 015012 (2003) [hep-ph/0209306].

\bibitem{crivellin}
 Crivellin A Kokulu A and Greub C 2013 Flavor-phenomenology of two-Higgs-doublet models with generic
Yukawa structure,"  Phys. Rev.D879, 094031 [arXiv:1303.5877 [hep-ph]]

\bibitem{36}
J. Adam et al. [MEG Collaboration], Phys. Rev. Lett. 110, no. 20, 201801 (2013) [arXiv:1303.0754 [hep-ex]].

\bibitem{37}
B. Aubert et al. [BaBar Collaboration], Phys. Rev. Lett. 104, 021802 (2010) [arXiv:0908.2381 [hep-ex]].

\bibitem{38}
K. Hayasaka et al. [Belle Collaboration], Phys. Lett. B 666, 16 (2008) [arXiv:0705.0650 [hep-ex]].

\bibitem{Dedes:2002er}
Dedes, Athanasios and Pilaftsis, Apostolos, Phys. Rev., D67,
2003, 
 015012,
10.1103/PhysRevD.67.015012.


 \bibitem{lito3lj}
 Y-F. Zhou, J.Phys.G30:783-792,2004.

 \bibitem{PDG}
  Olive, K. A. and others, Review of Particle Physics, Particle Data Group, Chin. Phys., C38, 2014.

\bibitem{Borzumati:1998tg}
  F.~Borzumati and C.~Greub,
  Phys.\ Rev.\ D {\bf 58}, 074004 (1998)
  [hep-ph/9802391].

\bibitem{Ciafaloni:1997un}
  P.~Ciafaloni, A.~Romanino and A.~Strumia,
  Nucl.\ Phys.\ B {\bf 524}, 361 (1998)
  [hep-ph/9710312].

\bibitem{Lees:2013uzd}
      Lees, J. P. et al.      
      Phys. Rev.D88, 2013, 7, 072012.
      
        
\bibitem{Crivellin:2012ye}
A. Crivellin, C. Greub and A. Kokulu, Phys. Rev. D 86,
054014 (2012) [arXiv:1206.2634 [hep-ph].

\bibitem{CHMhtaumu}
 V. Khachatryan et al. [CMS Collaboration], arXiv:1502.07400 [hep-ex].





\bibitem{EWC}
Alessandro Broggio, Eung Jin Chun, Massimo Passera, Ketan M. Patel, Sudhir K. Vempati, arXiv:1409.3199v1 [hep-ph].

\bibitem{Sola}
David Lopez-Val and Joan Sola, arXiv:1211.0311v3.




\end{thebibliography}
\end{document}